\documentclass[a4paper,11pt]{article}
\usepackage{jheppub} 
\usepackage{comment}
\usepackage{physics}
\usepackage{todonotes}
\usepackage{dsfont}
\usepackage{slashed}
\usepackage{hyperref}
\usepackage{url}
\usepackage{changepage}

\title{\begin{adjustwidth}{0cm}{-0.2cm} \boldmath \fontsize{19}{20}\selectfont Higgs-Induced Gravitational Waves: the Interplay of Non-Minimal Couplings, Kination and Top Quark Mass\end{adjustwidth}
}
\author[a]{Giorgio Laverda,}
\emailAdd{giorgio.laverda@tecnico.ulisboa.pt}
\author[b]{Javier Rubio}
\emailAdd{javier.rubio@ucm.es} 
\vspace{5cm}
\affiliation[a]{Centro de Astrof\'{\i}sica e Gravita\c c\~ao  - CENTRA,
Departamento de F\'{\i}sica, Instituto Superior T\'ecnico - IST,
Universidade de Lisboa - UL,
Av. Rovisco Pais 1, 1049-001 Lisboa, Portugal} 

\affiliation[b]{Departamento de Física Teórica and Instituto de Física de Partículas y del Cosmos (IPARCOS-UCM), Universidad Complutense de Madrid, 28040 
Madrid, Spain} 
\abstract{We explore a minimal scenario where the sole Standard-Model Higgs is responsible for reheating the Universe after inflation, produces a significant background of gravitational waves and maintains the full classical stability of the electroweak vacuum. As the Higgs self-coupling runs toward negative values at high energy scales, a non-minimal interaction with curvature during a stiff background expansion era drives the Higgs fluctuations closer to the instability scale. This curvature-induced tachyonic instability leads to an intense production of Higgs particles, accompanied by a stochastic gravitational-wave background. The characteristic features of such signal can be directly correlated to the inflationary scale, the non-minimal coupling parameter and the top-quark Yukawa coupling. We distinguish between three possible scenarios: absolute stability with low top-quark masses, potential vacuum instability, and absolute stability with new physics above the instability scale. Our findings suggest that the detection of a peaked background of gravitational waves together with its inflationary tail has the potential to unveil the features of the Higgs effective potential at very high energy scales while providing a minimal explanation for the reheating phase and the emergence of the Standard-Model plasma in the early Universe. Unlike other studies in the literature, the generation of gravitational waves in our scenario does not depend on the quantum instability of the Standard Model vacuum. }

\begin{document}

\maketitle
\flushbottom

\section{Introduction}

In recent years, the steady advancement of particle-collider experiments at the Large Hadron Collider has explored the Higgs sector up to energies of $13.6 \times 10^3\; \rm GeV$. The several orders of magnitude separating this energy range from plausible inflationary scales up to $10^{16} \textrm{ GeV}$ leave us wondering about new physical phenomena in the Higgs sector that might have taken place in the early Universe. What role does the Higgs play in the history of the early Universe? One possibility is that it dominates the energy budget at very high energy scales. This assumption identifies the Higgs field with the scalar degree of freedom responsible for inflation, giving rise to the Higgs inflation paradigm~\cite{Bezrukov:2007ep,Bauer:2008zj,Garcia-Bellido:2008ycs,Barbon:2009ya,Bezrukov:2009db,Bauer:2010jg,Burgess:2010zq,Bezrukov:2010jz,Giudice:2010ka,Bezrukov:2014bra,Hamada:2014wna,George:2015nza,Repond:2016sol,Fumagalli:2016lls,Bezrukov:2017dyv,Rubio:2019ypq,Shaposhnikov:2020fdv,Dux:2022kuk,Poisson:2023tja} (for a review see \cite{Rubio:2018ogq}). In the present work, we will consider the opposite alternative, namely that of a \emph{spectator} Higgs in the early Universe \cite{Markkanen:2017dlc, Markkanen:2017edu, Herranen:2014cua, Herranen:2015ima, Figueroa:2015rqa, Opferkuch:2019zbd, Laverda:2024qjt}. In this scenario, the Higgs field remains an energetically-subdominant quantum field during and shortly after inflation, with a negligible backreaction on the gravitational background. In general, the whole Standard Model (SM) sector coupled to the Higgs field can be considered to be in its vacuum state, while the Universe is still far from achieving a thermalised state. 

In this context, non-minimal gravitational interactions can induce substantial new-physics effects within the Higgs sector. Studies of quantum fields in curved spacetime reveal in fact that tree-level operators associated with these interactions are naturally generated via the renormalisation of the energy-momentum tensor \cite{Birrell:1982ix}. Assuming the corresponding coupling constants to vanish is therefore an ad hoc prescription valid only at a specific energy scale, since such a choice is not preserved under radiative corrections as first noted in \cite{Chernikov:1968zm,Callan:1970ze,Tagirov:1972vv}.
With these considerations, the physics of the Higgs field in the early Universe is dictated by an effective Lagrangian density
\begin{equation} \label{eq:higgs_lagrangian}
    \mathcal{L}= \left( \frac{M^2_{P}}{2} - \xi H^{\dagger}H  \right) R - g^{\mu \nu} (D_{\mu}H)^{\dagger}(D_{\nu}H) - \lambda \left( H^{\dagger}H - \frac{v^2_{_{\hspace{-0.03cm} \textrm{EW}}}}{2} \right)^2  + \mathcal{L}_{\rm SM} + \mathcal{L}_{\rm BSM} \,,
\end{equation}
with the potential for the Higgs doublet $H$ depending on the quartic self-coupling $\lambda$ and the electroweak vacuum expectation value $v_{_{\hspace{-0.03cm} \textrm{EW}}}$. The additional term $\mathcal{L}_{\rm SM}$ encompasses all SM interactions beside the Higgs-field kinetic and potential terms, while $\mathcal{L}_{\rm BSM}$ accounts for contributions from Beyond the Standard-Model (BSM) physics at high energy scales. In particular, $\mathcal{L}_{\rm BSM}$ includes the kinetic and potential terms of an inflationary degree of freedom $\phi$, which is assumed to dominate the energy content of the Universe during its earliest stages of evolution. In this work, we will remain agnostic about the specific inflationary framework governing the spectator Higgs field dynamics. Our only requirement is that inflation is followed by a kination or stiff expansion era, with the end of slow-roll resulting in a kinetically-dominated Universe with effective equation-of-state parameter ${w_{\phi}=p_{\phi}/\rho_{\phi}\simeq 1}$. Such a sequence of background expansion phases can be found typically in models of \emph{quintessential inflation}, where a steep potential interpolates between two plateaus at high and low scales in order to describe the inflationary and cosmological-constant epochs with a single scalar degree of freedom \cite{Peebles:1998qn,Spokoiny:1993kt, Bettoni:2021qfs}. The subdominant Higgs field is assumed to be coupled only indirectly through gravitational interactions to the inflationary sector, with suppressed tree-level interactions in the Lagrangian \eqref{eq:higgs_lagrangian}. These features can be readily achieved by assuming a shift-symmetric inflaton field, such as in the case of variable gravity settings \cite{Wetterich:1987fm,Wetterich:1994bg,Rubio:2017gty,Wetterich:2019qzx}, which gives rise to an almost massless inflaton in the asymptotic regimes. However, other scenarios can include a inflation-to-kination transition as well, including asymmetric $\alpha$-attractors models \cite{Akrami:2017cir, Dimopoulos:2017zvq,Dimopoulos:2017tud, Garcia-Garcia:2018hlc}, and axion-like models \cite{Gouttenoire:2021jhk}. 

The transition from inflation to kination triggers a so-called Hubble-Induced Phase Transition (HIPT) \cite{Laverda:2023uqv, Laverda:2024qjt, Bettoni:2019dcw, Bettoni:2021zhq, Bettoni:2018utf, Bettoni:2018pbl, Mantziris:2024uzz, Kierkla:2023uzo}. More precisely, in a Friedmann--Lema\^itre--Robertson--Walker (FLRW) spacetime with metric tensor $g_{\mu\nu}={\rm diag}(-1, \, a^2(t)\delta_{ij})$, the gravitational contribution to the Higgs mass is determined by the Ricci or curvature scalar 
\begin{equation}
    R=3(1-3w)\mathcal{H}^2 \; ,
\end{equation}
with $\mathcal{H}=\dot a/a$ the Hubble function and $w$ the global equation-of-state parameter, which, within the spectator field approximation, can be safely identified with that of the inflation field, $w\simeq w_{\phi}$. A variation of the background expansion rate of the Universe impacts therefore the Higgs effective mass, making it possible for new phases to appear in its effective potential. In particular, for inflationary scenarios involving a stiff expansion era following the end of inflation, the curvature scalar experiences a transition from positive ($w=-1$) to negative values ($w=1$).

During inflation, the large Hubble-induced mass suppresses the Higgs quantum fluctuations, both at super-horizon and sub-horizon scales, while preventing the generation of large isocurvature fluctuations \cite{Laverda:2024qjt, Herranen:2014cua, Herranen:2015ima, Kohri:2016wof}. At the onset of kination, however, the large negative gravitational mass term induces a spontaneous symmetry-breaking with Hubble-dependent true vacua. The Higgs quantum modes then experience an exponential tachyonic amplification that quickly builds up the occupation number of infrared (IR) modes, followed by a cascade effect that turbulently populates ultraviolet (UV) modes as well.~\footnote{It is worth noting that this dynamic is significantly different from the well-known scenario of a resonantly-excited Higgs field interacting with an oscillating background \cite{Figueroa:2015rqa, Figueroa:2016dsc, Figueroa:2017slm, Figueroa:2016wxr, Herranen:2015ima, Mantziris:2020rzh, Mantziris:2021oah, Mantziris:2021zox, Mantziris:2022bfe, Mantziris:2022fuu, Mantziris:2023xsp, Kohri:2016wof, Postma:2017hbk, Ema:2017loe, Enqvist:2016mqj, Ema:2016kpf}.} It is precisely these amplified fluctuations that provide us with a way to study the post-inflationary spectator Higgs as they lead to three main outcomes:
\begin{enumerate}

    \item The interplay between kination and non-perturbative tachyonic particle production allows the spectator Higgs field itself to (re)heat the Universe, even without direct couplings of the inflaton field to matter or a full depletion of the inflaton condensate. Requiring the Universe to be radiation-dominated at temperatures higher than the Big-Bang-Nucleosynthesis (BBN) temperature \cite{deSalas:2015glj, Hasegawa:2019jsa} sets constraints on the space of model parameters.  
    
    \item Fluctuations that are significantly amplified can potentially probe the dangerous region above the instability scale $\mu_{\rm inst}$ \cite{Tang:2013bz, Espinosa:2015qea}, defined by the Higgs self-coupling running to negative values, i.e. $\lambda(\mu_{\rm inst})=0$. Requiring the classical stability of our Universe imposes constraints on the top-quark mass, as this is the major actor in the negative running of $\lambda$ \cite{Bezrukov:2014bra, Hiller:2024zjp}.

    \item The generation of large spatial gradients leads inevitably to the production of a Stochastic Gravitational-Wave Background (SGWB). The signal is generated as the classicalised Higgs fluctuations enter the post-tachyonic non-linear dynamics, when the self-scattering process fragments the homogeneous patches generated by the tachyonic instability inside the Hubble volume \cite{Bettoni:2024ixe}.
    
\end{enumerate}

\begin{figure}[tb]
\centering
\includegraphics[width=0.85\textwidth]{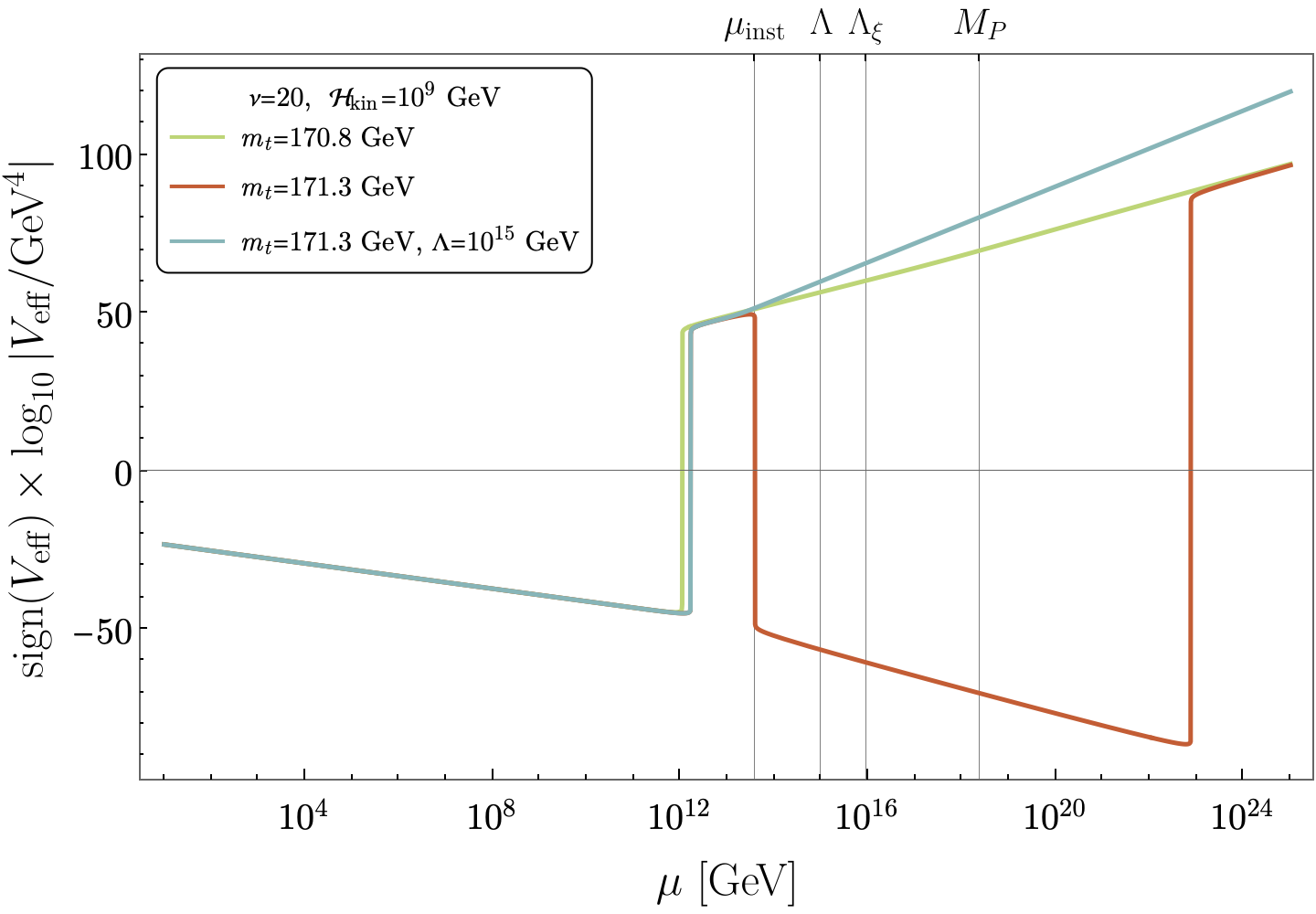}
\hspace{5mm}
\caption{Higgs effective potential with a non-minimal gravitational coupling at the onset of kination for three prototypical scenarios: absolute stability (green line, $m_t=170.8 \textrm{ GeV}$), instability (red line, $m_t=171.3 \textrm{ GeV}$), stability with new physics as sextic operators at a typical scale ${\Lambda=10^{15} \textrm{ GeV}}$ (blue line, $m_t=171.3 \textrm{ GeV}$). The cut-off of the effective gravitational theory is given by $\Lambda_{\xi}=9.2\times10^{15} \textrm{ GeV}$. The cosmological parameters have been set to $\nu=20$ and $\mathcal{H}_{\rm kin}=10^{9} \textrm{ GeV}$.}\label{fig:effective_potential}
\end{figure}

The SGWB signal generated by HIPTs offers a unique window into the post-inflationary physics of the spectator Higgs field. In particular, the specific features of the Gravitational-Wave (GW) spectrum allow us to explore the running of the Higgs self-coupling at high energy scales, by indirectly measuring the top-quark mass. Three possible scenarios can occur. For top-quark masses below the critical value of $170.886 \textrm{ GeV}$ that guarantees absolute stability, the GW signal can peak at $h^2\Omega_{\rm GW,0}\sim10^{-9}$ and its typical frequency ${f_{\rm GW,0}\sim10^{10} \textrm{ Hz}}$ is tightly linked to the top-quark mass. For higher top-quark masses, the constraint on the stability of the electroweak vacuum bounds the inflationary scale from above, thus limiting the amplitude of the signal to $h^2\Omega_{\rm GW,0}\lesssim10^{-9}$. Notably, this GW background does not rely on the Higgs field crossing the instability scale, as considered, for instance, in \cite{Espinosa:2018eve}.  In the presence of new physics around the instability scale, a prototypical sextic operator in the Higgs potential can restore its absolute stability, see Figure \ref{fig:effective_potential}. In this case, a high-scale phase transition is still compatible with stability even for high inflationary scales or values of the top-quark mass above the critical one.

Exploring a large range of non-minimal coupling parameters $\xi$, inflationary scales $\mathcal{H}_{\rm inf}$, and top-quark masses $m_{t}$, allows us to identify telltale features in the GW spectrum. Our semi-analytical approach makes use of parametric formulas derived in previous works that condense numerous fully-fledged 3+1 classical lattice simulations into a few simple expressions \cite{Laverda:2023uqv, Bettoni:2024ixe}. We also perform three high-resolution benchmark simulations in the scenario involving non-renormalisable operators to check the implications for the GW signal. 

This work is structured as follows. Section \ref{sec:spectatorHiggs} provides an overview of the Higgs field's behaviour during a HIPT transition, with particular focus on the validity of the spectator field approximation and the consistency of the effective field theory approach. In Section \ref{sec:eff_pot}, we examine the impact of radiative corrections, emphasising the connection between the key cosmological parameters governing (re)heating dynamics and low-energy Standard Model inputs such as the top-quark mass. Section \ref{sec:GW_signal} then presents a general analysis of the gravitational wave spectrum produced by HIPTs, which is further applied to the Standard Model Higgs scenario in Section \ref{sec:probing_top_mass}. This discussion highlights the differences between cases of absolute stability, instability, and an enhanced (or independent) heating sector. In Section \ref{sec:bsm_stability}, we explore the stabilisation of the effective potential by new physics at high energy scales. Finally, Section \ref{sec:conclusion} offers concluding remarks and outlines directions for future research. 

\section{The spectator Higgs field approximation} \label{sec:spectatorHiggs}

To qualitatively discuss the general behaviour of the Higgs field during a HIPT \cite{Bettoni:2019dcw, Laverda:2023uqv, Bettoni:2021zhq} and assess the validity of the spectator field approximation, it suffices to focus on the portion of \eqref{eq:higgs_lagrangian} that involves the inflaton field $\phi$ and the subdominant Higgs field in the unitary gauge $H=(0,\,h/\sqrt{2})^T$, namely 
\begin{equation}
 {\cal L} =\frac{M^2_P}{2}R + \mathcal{L}_{\phi}  -\frac12{\partial^{\mu}}{h} \partial_{\mu} h - \frac12 \xi R h^2 - \frac{\lambda}{4}h^4 \,,    \label{eq.lagrangian_chi_scalar}
\end{equation}
 where we have intentionally omitted the Higgs vacuum expectation value $v_{_{\hspace{-0.03cm} \textrm{EW}}}$, as this will not play any essential role at the high energy scales considered in this paper. The dynamics of the Higgs field is governed by the Klein--Gordon equation,
\begin{equation}\label{eq:eq_hcov}
 \Box\, h -\xi R h -\lambda h^3=0\,,
\end{equation}
along with the energy density and pressure following from the $00$ and $ii$ components of the associated energy-momentum tensor,
\begin{equation}
T^{(h)}_{\mu\nu}  = 
  \partial_{\mu} h \partial_{\nu} h - g_{\mu\nu} \left[
   \frac{1}{2} \partial^\rho h \partial_\rho h    
  + \frac{\lambda}{4} \, h^4 \right]  +\xi \Big[ G_{\mu\nu} 
  + \left(g_{\mu\nu} \Box -\nabla_{\mu} \nabla_{\nu}\right)
 \Big] h^2,    
  \label{eq:emt_h}        
\end{equation}
with $\nabla_\mu$ the covariant derivative, $G_{\mu\nu}$ the Einstein tensor, and $\Box = \nabla^{\lambda} \nabla_{\lambda}$ defines the d'Alambertian operator. Notably, the contraction of $T^{(h)}_{\mu\nu}$ with arbitrary timelike vectors $u^\mu$ and $u^\nu$ is not necessarily positive definite at all times \cite{Ford:1987de,Ford:2000xg,Bekenstein:1975ww,Flanagan:1996gw},  leading to a violation of the weak energy condition $T^{(h)}_{\mu\nu} u^\mu u^\nu \geq 0$ within this particular sector. 

The transition from an inflationary epoch to a kination-dominated phase induces a significant tachyonic mass in the Higgs effective potential, triggering a second-order phase transition at high-energy scales. To determine the limit of validity of the spectator field approximation, we examine the Einstein equations 
\begin{equation}
M_P^2 \, G_{\mu\nu}=T^{(h)}_{\mu\nu}+T^{(\phi)}_{\mu\nu}\,,
\end{equation}
with $ T^{(\phi)}_{\mu\nu}$ the inflaton energy-momentum tensor. Taking the trace of this expression, performing a spatial average of both sides of the result, and using \eqref{eq:emt_h} to solve explicitly for $R$, we obtain 
\begin{equation}
M_P^2 \, R = \langle T^{(\phi)}\rangle   \frac{1-\Delta_1}{1+\Delta_2}\,,
\end{equation}
with 
\begin{equation}
\Delta_1=(6\xi-1)\frac{ \langle T_{\xi=0}^{(h)}\rangle}{\langle T^{(\phi)}\rangle}\,, \quad \quad \quad  \Delta_2=(6\xi-1)\frac{\xi\langle h^2\rangle}{M_P^2}\,, 
\end{equation}
and $\langle T^{(\phi)}\rangle$ and  $\langle T_{\xi=0}^{(h)}\rangle$ the average traces of the inflaton energy-momentum tensor and the Higgs energy-momentum tensor \eqref{eq:emt_h} \textit{at zero non-minimal coupling}. In the conformal limit ($\xi = 1/6$) and whenever the corrections $\Delta_1$ and $\Delta_2$ are negligible, the evolution of the Ricci scalar is entirely determined by the homogeneous inflaton component. These two corrections, however, play distinct roles. The term $\Delta_1$ controls the ratio between the $\xi$-enhanced energy density of the spectator field and that of the inflaton, thereby affecting potential sign changes in the Ricci scalar and ultimately governing the termination of the tachyonic instability. For sufficiently small self-couplings, this condition is equivalent to imposing the constraint $\langle h^2 \rangle^{1/2} \ll M_P/\sqrt{\xi}$
\cite{Figueroa:2024asq}, in full agreement with the assumptions in previous studies \cite{Bettoni:2019dcw, Laverda:2023uqv, Bettoni:2021zhq}.
Meanwhile, $\Delta_2$ quantifies the degree of non-linearity introduced by the non-minimal coupling of the Higgs field to gravity, in contrast to its behaviour in the minimally coupled Standard Model and General Relativity. As shown in Appendix \ref{app:cut-off}, the condition $\Delta_2 \ll 1$ ensures that the typical fluctuations of the Higgs field remain well below the quantum cut-off scale $
\Lambda_\xi = M_P/\xi$ of this inherently non-renormalizable framework \cite{Barbon:2009ya,Burgess:2009ea,Hertzberg:2010dc,Burgess:2010zq,Bezrukov:2010jz,Bezrukov:2012hx,Ren:2014sya,Bezrukov:2014ipa,Fumagalli:2017cdo,Antoniadis:2021axu,Mikura:2021clt,Ito:2021ssc,Escriva:2016cwl,Karananas:2022byw}. Beyond this threshold, and in accordance with the standard effective field theory logic, the starting Lagrangian density \eqref{eq:higgs_lagrangian} must necessarily be supplemented with an infinite series of higher-dimensional operators to maintain consistency, \textit{regardless of the presence or absence of backreaction on the inflaton field}.

\section{Electroweak vacuum stability and the onset of heating} \label{sec:eff_pot}

Due to the absence of a global minimum, runaway inflationary potentials like the one under consideration do not lead to standard reheating scenarios based on non-perturbative resonant production. On top of that, the suppression of tree-level interactions between the Higgs and the inflaton fields in the Einstein frame does not allow for reheating mechanisms à la instant reheating \cite{Felder:1998vq}. Indeed, within the validity regime of the graviscalar effective field theory, a possible interaction portal between the inflaton and the Higgs field is not active, since the Einstein-frame Lagrangian contains only Planck-suppressed tree-level interactions between the two, see Appendix \ref{app:cut-off}. 
We also assume that, if a direct tree-level coupling of the form $\sim g^2 h^2 \phi^2$ is present in the Jordan frame—explicitly breaking the advocated shift symmetry of the inflationary potential near the crossover scale—it is suppressed by a small coupling $g$. \footnote{This coupling causes a violation of the adiabaticity condition in the spectator field and the production of particle excitations. However, the efficiency of such process is proportional to $g^2$ and therefore suppressed for perturbative couplings. For instance, setting $\mathcal{H}_{\rm kin}=10^{11} \textrm{ GeV}$ and $\lambda\sim10^{-2}$, this this contribution is subdominant with respect to the tachyonic production over the whole parameter space under consideration already for $g\sim10^{-5}$ \cite{Bettoni:2021qfs, Laverda:2023uqv}.}

As a consequence, leveraging on the time-dependence of the Higgs effective mass, the tachyonic amplification of Higgs fluctuations during HIPTs offers a compelling mechanism for heating the Universe in non-oscillatory models of inflation. This amplification is typically a brief process, as both backreaction effects and the rapid decrease of the Hubble rate cause the non-minimal interaction to quickly become subdominant, effectively restoring the original symmetry of the potential \cite{Bettoni:2019dcw,Bettoni:2021zhq,Laverda:2023uqv}. Due to the quartic shape of the SM Higgs potential, the classicalised Higgs energy density evolves asymptotically as radiation, \(\rho_h \sim a^{-4}\), allowing it to grow over the rapidly decaying inflaton background, which, for a stiff equation of state $w\simeq w_\phi\simeq 1$, evolves as \(\rho_\phi \sim a^{-6}\). The final viability of this appealing heating scenario is nonetheless deeply connected to the stability of the Standard Model vacuum at high energies. To see this, let us consider the  zero-temperature Renormalisation-Group-Improved (RGI) potential for the non-minimally coupled Higgs field, 
\begin{equation}\label{eq:effHiggspot}
    V_{\rm eff}(h) = \frac12 \xi R h^2 + \frac14 \lambda(\mu)  h^4\,,
\end{equation}
with the scale $\mu$ in the running coupling constant $\lambda(\mu)$ set as a combination of the Higgs field itself and the cosmological background scale, $\mu^2=\mathcal{H}^2+h^2$, as per the standard prescription on curved cosmological backgrounds \cite{Markkanen:2018bfx, Markkanen:2017dlc}.

To facilitate the analytical and numerical treatments, we start by performing a fitting of the full numerical three-loop $\overline{\rm MS}$ running of the Higgs self coupling \cite{BezrukovNotebook} parametrising its dependence on the top quark mass. Motivated by their small experimental error, we set the Higgs mass to the central value of the current best measure $m_h=125.20\pm0.11 \textrm{ GeV}$ and the strong coupling constant to $\alpha_s=0.1180$ \cite{ParticleDataGroup:2024cfk}. A good ansatz in the energy-scale range $\mu\in[10^4 \textrm{ GeV}, \, 10^{23} \textrm{ GeV}]$ for the parametric shape is given by
\begin{equation}
    \lambda_{\rm fit}(\mu)=\lambda_0+c_1\log \left( \frac{\mu}{\rm GeV} \right) +c_2\log^2 \left( \frac{\mu}{\rm GeV} \right) +c_3\log \left( \log\left(\frac{\mu}{\rm GeV}\right) \right)
    \label{eq:running_fit}
\end{equation}
where the top-mass-dependent coefficients $\lambda_0$, $c_1$, $c_2$ and $c_3$ are given by
\begin{align}
    \lambda_0(m_t)&=7.31965 -9.13428\times 10^{-2}\times \frac{m_t}{\rm GeV}+2.94355\times 10^{-4}\times \left(\frac{m_t}{\rm GeV}\right)^2 \; ,\nonumber \\
    c_1(m_t)&=1.98329\times 10^{-1} -2.34181\times 10^{-3}\times \frac{m_t}{\rm GeV}+7.05298\times 10^{-6}\times \left(\frac{m_t}{\rm GeV}\right)^2 \; ,\nonumber \\
    c_2(m_t)&=-1.4835\times 10^{-4} + 7.88043\times 10^{-7}\times \frac{m_t}{\rm GeV} \; ,\nonumber \\
    c_3(m_t)&=-4.19967 + 5.30679\times 10^{-2}\times \frac{m_t}{\rm GeV}-1.70896\times 10^{-4}\times \left(\frac{m_t}{\rm GeV}\right)^2 \; .
\end{align}
The accuracy of such parametric function is always within 1\% of  the full numerical running for the renormalisation scales under considerations and for a wide range of top quark masses $m_t\in[169 \textrm{ GeV} \; ,175 \textrm{ GeV}]$.
Note that in these expressions, $m_t$ refers to the top-quark \textit{pole} mass  $m_t^{\text{pole}}$, which differs from the \textit{Monte Carlo reconstructed} mass $m_t^{\text{rec}}$ typically obtained using event generators such as PYTHIA or HERWIG \cite{Hoang:2020iah, Nason:2017cxd}. The discrepancy between these two scales is approximately 1--2 GeV, with recent measurements consistently finding $m_t^{\text{pole}}$ to be lower than $m_t^{\text{rec}}$. Specifically, $m_t^{\text{pole}} = 170.5 \pm 0.8$ GeV~\cite{CMS:2019esx}, while $m_t^{\text{rec}} = 171.77 \pm 0.37$ GeV~\cite{CMS:2023ebf}. 

A graphical representation of the effective potential \eqref{eq:effHiggspot} can be seen in Figure \ref{fig:effective_potential}, where the green and red lines exemplify the possible formation of a barrier in the SM effective potential at the instability scale $\mu_{\rm inst}$ for different choices of top-quark pole mass. Beyond that scale, the Higgs potential becomes negative, with the new and significantly deeper minimum appearing at super-Planckian field values in the absence of new physics.

Crossing the electroweak vacuum instability threshold after a HIPT could trigger the formation of anti-de Sitter regions expanding at the speed of light and ultimately engulfing our observable Universe \cite{Espinosa:2015qea}. To prevent this catastrophic scenario, a stability constraint must be imposed on the energy density of tachyonically amplified Higgs fluctuations. The parametric description derived in a previous study from an extensive set of numerical 3+1 classical lattice simulations \cite{Laverda:2023uqv} provides a practical method for estimating the Higgs energy density when backreaction effects become relevant. Indeed, lattice simulations allow for the evolution of the interacting Higgs field in a high-occupation-number regime while fully incorporating the backreaction arising from self-interactions in the fundamental Lagrangian. By examining the complete dynamics, one can define a backreaction timescale, i.e. the moment when Higgs fluctuations reach large amplitudes and self-scattering effects become significant. The energy density at this backreaction time can then be compared to the height of the barrier in the effective potential, thereby establishing a stability constraint,
\begin{equation} \label{eq:barrier_overcoming}
   \rho_{\text{tac}}(\mathcal{H}_{\rm kin},\nu, \lambda(h_{\rm min})) < V_{\rm eff}^{\rm 3 loop}(h_{\rm max}) \;,
\end{equation}
with $\rho_{\rm tac}$ the Higgs energy density including the non-minimal coupling terms at the end of the tachyonic phase. The field values $h_{\rm min}$ and $h_{\rm max}$ correspond to the positions of the new Hubble-induced minimum and the top of the barrier, found by solving $\partial V_{\rm eff} / \partial h=0$. In order to obtain a good estimate of the Higgs self-coupling at the time when backreaction effects set in, we compute $\lambda(h_{\rm min})$ at the time of backreaction $z_{\rm br}\sim10$, which corresponds to $\mathcal{H}_{\rm br}\sim 0.5\mathcal{H}_{\rm kin}$ \cite{Opferkuch:2019zbd}. Indeed, the top-mass-dependent $h_{\rm min}$ scale is the typical amplitude of Higgs fluctuations a the end of the tachyonic phase.

Our numerical analysis \cite{Laverda:2023uqv} using 3+1-dimensional numerical simulations with the code \texttt{$\mathcal{C}osmo\mathcal{L}attice$} \cite{Figueroa:2021yhd, Figueroa:2016wxr} found the following parametric behaviour for the energy-density of the Higgs field at the end of the tachyonic phase
\begin{equation}
    \rho_{\text{tac}}(\mathcal{H}_{\rm kin}, \nu, m_t) = 16 \, \mathcal{H}^4_{\rm kin}\, \exp\left(\beta_1(m_t) + \beta_2(m_t) \, \nu+ \beta_3(m_t) \log \nu \right)  \,.
\label{eq:fit_rho_br}
\end{equation} 
The coefficients $\beta_1,\beta_2$ and $\beta_3$ in this expression are given in terms of $n=-\log_{10}(\lambda(\mu))$, but the running in \eqref{eq:running_fit} allows us to trade the dependence on $\lambda$ for a dependence on $m_t$, thus making explicit the connection between cosmological and SM parameters. The quantity $\mathcal{H}_{\rm kin}$ in this expression is the Hubble scale corresponding to the beginning of the kination phase which, for an instantaneous transition like the one under consideration, \footnote{One might wonder if taking into account the non-instantaneous transition from inflation to kination can lead to a non-negligible contribution to the Higgs energy density. This is important whenever the tachyonic process is inefficient, i.e. when the non-minimal-coupling induced vacuum is displaced at small field values. In general, a scalar field with a time-varying effective mass between inflation and kination undergoes a process of non-adiabatic particle production. This can lead to some initial excitation that can contribute to the final energy-density after the Hubble-induced phase transition. One can estimate the amplitude reached after this non-adiabatic excitation following the references \cite{Mukhanov:2007zz, Herranen:2015ima}, and find it to be $\langle h^2 \rangle \sim \mathcal{H}^2_{\rm kin} \xi $. For Higgs-like values of $\lambda(\mu)$, one concludes that the tachyonic amplification is more effective by at least one order of magnitude. Modelling the non-instantaneous onset of kination has been considered in \cite{Figueroa:2024asq} by including an inflaton field together with a specific potential that defines a model-dependent transition.} can be identified with the inflationary scale $\mathcal{H}_{\rm kin} \approx \mathcal{H}_{\rm inf}$. A detailed discussion of the stability/instability scenarios was carried out in \cite{Laverda:2024qjt}, alongside the study of the heating stage in the Higgs, gauge and fermionic sectors, concluding than only the former is relevant for the macroscopic dynamics of heating.\footnote{Resonant production of gauge bosons occurs as the Higgs oscillates in its effective potential. While this process involves successive exponential amplification, its efficiency is limited compared to the rapid tachyonic production of the Higgs itself. Assuming the Higgs oscillates uniformly in a quartic potential, resonant amplification proceeds with a Floquet index of $\mu_k \to \mu_{\text{max}} = 0.2377$ for $q = g^2/\lambda \gg 1$, typical for SM couplings \cite{Greene:1997fu}. Even under optimal conditions, it takes at least 10 oscillations for the energy density of bosonic fields to match that of the Higgs. The additional decay of gauge bosons into fermions does not significantly impact the heating timeline or vacuum stability constraints. However, non-Abelian interactions may aid in the thermalisation of the SM plasma, accelerating the approach to equipartition through boson scatterings and annihilations \cite{Enqvist:2015sua,Bodeker:2007fw,Bezrukov:2014ipa,Repond:2016sol}.} Here we only emphasise that the same set of parametric formulas derived in \cite{Laverda:2023uqv} allows us to quantify the course of the heating stage. Indeed, the first consequence of the Hubble-induced tachyonic instability is the enhancement of the Higgs energy density and the heating of the primordial Universe. To describe this phase, we make use of the \emph{radiation} timescale at which the equation-of-state parameter for the Higgs field becomes $w_h=1/3$, 
\begin{equation}
    z_{\text{rad}}(\nu, m_t) = \gamma_1(m_t) + \gamma_2(m_t) \, \nu \,, 
\label{eq:fit_z_rad}
\end{equation}   
and of the Higgs energy density
\begin{equation}
    \rho_{\text{rad}}(\mathcal{H}_{\rm kin}, \nu, m_t) = 16 \mathcal{H}^4_{\rm kin} \, \exp\left({\delta_1(m_t) + \delta_2(m_t) \, \nu} +\delta_3(m_t)\log \nu \right) \,, 
\label{eq:fit_rho_rad}
\end{equation}  
evaluated at $z_{\rm rad}$, with $z = a_{\rm kin}\sqrt{6\xi}\mathcal{H}_{\rm kin}\tau$ a rescaled conformal-time variable. The full expressions, including the numerical coefficients, can be found in Appendix \ref{app:parametric_formulas}. The duration of the heating phase is effectively encoded in the so-called \emph{heating efficiency} of the Higgs sector \cite{Rubio:2015zia},
\begin{equation} \label{eq:heating_eff_def}
    \Theta_{\text{ht}}^{h} \equiv \frac{\rho_{h}(a_{\rm rad})}{\rho_{\phi}(a_{\rm rad})} \,,
\end{equation}
which measures the intensity of the non-perturbative tachyonic particle production. We also associate a \textit{radiation temperature} to the Higgs-heated Universe \cite{Rubio:2017gty}
\begin{equation}\label{eq:temp_reheating}
T_{\rm ht} =\left(\frac{30\,\rho^{\rm ht}_{h}}{\pi^2 g_*^{\rm ht}}\right)^{1/4} \,,   
\end{equation}
with $g_*^{\rm ht}=106.75$ the SM number of relativistic degrees of freedom at energies above ${\cal O}(100) \text{ GeV}$ and $\rho^{\rm ht}_{h}=\rho^{\rm ht}_{\phi}$ the total energy density of the Higgs field at the end of the heating phase. The fitting formula for $\rho_{\rm rad}$ in \eqref{eq:fit_rho_rad} allow us to estimate the (re)heating temperature in terms of the model parameters $(\mathcal{H}_{\rm kin}, \, \nu, \,  m_t)$. Note that our definition of radiation temperature is agnostic about the achievement of a thermalised Universe. However, the thermalisation timescale is typically similar to the duration of the heating phase \cite{Micha:2004bv, Bettoni:2021zhq} and we will assume the Universe to be thermalised at the end of the heating stage.

If the breaking scale is sufficiently high $\mathcal{H}_{\rm kin} \gtrsim 10^{5.5} \text{ GeV}$ \cite{Laverda:2024qjt}, the Higgs heating efficiency is large enough to dominate the heating stage with the constraint ${T_{\rm ht}>5 \text{ MeV}}$ \cite{deSalas:2015glj, Hasegawa:2019jsa} generally fulfilled. In other words, the Higgs alone possesses the required energy density to start a phase of radiation-domination before BBN and no other heating mechanism is required. The parametric formulas in \cite{Laverda:2023uqv} can be used directly to compute the heating efficiency
\begin{equation} \label{eq:fittingTheta}
    \Theta_{\text{ht}}^{h}(\mathcal{H}_{\rm kin},\nu, m_t) = \frac{\rho_{\rm rad}(\mathcal{H}_{\rm kin},\nu, m_t)}{3 \mathcal{H}_{\rm kin}^2 M_{P}^2}  \left(1+\frac{z_{\rm rad}}{\nu}\right)^3  \,,
\end{equation}
and the (re)heating temperature
\begin{eqnarray} \label{eq:reheating_temp}
    T_{\text{ht}} &\simeq& 2.7 \times 10^{8} \,\text{GeV} \left(\frac{a_{\rm rad}}{a_{\rm kin}}\right)^{-3/2} \left( \frac{\Theta_{\text{ht}}^{h}}{10^{-8}}\right)^{3/4}  \left( \frac{\mathcal{H}_{\rm kin}}{10^{11} \, \text{GeV}}\right)^{1/2}  \,.
\end{eqnarray}
The successful heating of the Universe depends mostly on the scale of kination and on the non-minimal coupling parameter $\nu$, making the threshold scale $\mathcal{H}_{\rm kin} \simeq 10^{5.5} \text{ GeV}$ independent from the chosen value of $m_t$. Indeed, for energy scales far below the instability scale, the running of the Higgs self-coupling converges to $\lambda(\mu)\simeq 0.1$ for all top-quark masses, and only a weak dependence on $\nu$ remains for small values of non-minimal interactions.

\section{Stochastic gravitational-wave background} \label{sec:GW_signal}

The second major consequence of the initial amplification of the Higgs quantum modes is the generation of large fluctuations on sub-horizon scales. These ``bubbly" features in the field distribution lead to the appearance of large spatial gradients \cite{Laverda:2023uqv}, thus sourcing a strong GW signal \cite{GarciaBellido:2008ab}. This production phase is limited to the interval between backreaction and the radiation timescale $z_{\rm rad}$, during which the non-minimal and self-interaction contributions have comparable magnitudes. By the radiation time, the system has quenched into a state of fragmented, small fluctuations. Hence, the GW spectrum possesses a clear peaked shape, as is typical for transient sources. The features of such GW signal were studied in detail with 3+1 classical lattice simulations in a recent work \cite{Bettoni:2024ixe}. The general shape of the spectrum is well described by a prototypical broken-power-law shape function \cite{Lewicki:2020azd, Lewicki:2022pdb}
\begin{equation}
    \Bar{\Omega}_{\rm GW}(\mathcal{H}_{\rm kin},\nu, m_t; f)=\frac{\Bar{\Omega}_{\rm p}(a+b)^c}{\left[a \left(\frac{f}{f_{\rm p}}\right)^{b/c}+b \left(\frac{f}{f_{\rm p}}\right)^{-a/c}\right]^c}\,,
\end{equation}
where the momentum and amplitude at the peak are given in terms of parametric functions obtained from numerical 3+1 classical lattice simulations,
\begin{gather}
    \kappa_{\rm p}(\nu, m_t) = \alpha_1(m_t) + \alpha_2(m_t) \nu \; , \nonumber \\ 
    \Bar\Omega_{\rm GW, p}(\mathcal{H}_{\rm kin},\nu, m_t) = \left(\frac{\mathcal{H}_{\rm kin}}{10^{10} \textrm{ GeV}}\right)^2 \exp \left[ \beta_1 (m_t) + \beta_2 (m_t) \log \nu \right] \; ,
    \label{eq:fit_momentum_peak}
\end{gather}
with momenta $\kappa$ normalised with respect to $\mathcal{H}_{\rm kin}$. \footnote{Note that the typical momenta around the peak of the GW spectrum are typically $ \sim 10-100 \times \mathcal{H}_{\rm kin}$ and are well below the effective-field-theory cut-off scale for $\mathcal{H}_{\rm kin}\lesssim 10^{12} \textrm{ GeV}$. } As noted in the previous section, the dependence on the running of the Higgs self-coupling has been recast into a dependence on the top-quark mass, thanks to the analytic form of the running in \eqref{eq:running_fit}. The IR slope of the spectrum follows the expected causal behaviour $ a = 3.00 $ \cite{Caprini:2009fx, Cai:2019cdl}, while the UV slope depends on the dynamics of the turbulent cascade following the out-of-equilibrium tachyonic particle production with typical values of the parameters $b = 152.34 - 6.57 \, \nu$ and $c=105.85 - 4.79 \, \nu$. Spectra-related quantities are evaluated at the radiation time, i.e. at the end of the violent GW production process. Moreover, we have used the following normalisation of the GW energy density
\begin{equation}
    \Bar\Omega_{\rm GW}(z_{\rm rad})\equiv\frac{\rho_{\rm GW}(z_{\rm rad})}{\rho_{h}(z_{\rm rad})} \,,
\end{equation}
with the critical density identified with the total energy density in the Higgs sector. Present-time frequencies and energy-densities can be computed using the expressions \cite{Allahverdi_2021} 
\begin{gather}
    f_{\rm GW,0} (\nu, m_t)\simeq 1.3\times 10^9 \, \text{Hz} \; \frac{\kappa}{2\pi} \left(\frac{\mathcal{H}_{\rm kin} \, a_{\rm rad}}{10^{10} \textrm{ GeV}}\right)^{1/2} \left( \frac{\Theta_{\rm ht}^h}{10^{-8}}\right)^{-1/4} \,, \nonumber \\
    \Omega_{\rm GW,0} (\mathcal{H}_{\rm kin},\nu, m_t) = 1.67 \times 10^{-5} h^{-2} \left( \frac{100}{g_*^{\rm ht}} \right)^{1/3} \times \Bar{\Omega}_{\rm GW}\,,
    \label{eq:f_0}
\end{gather}
which depend on the duration of kination via the heating efficiency $\Theta_{\rm ht}^h$. As discussed in \cite{Bettoni:2024ixe}, the typical frequency of the spectrum does not depend on the scale of kination because of the cancellation caused by the common dependence on $\mathcal{H}_{\rm kin}$ of all relevant quantities. The heating efficiency $\Theta_{\rm ht}^h$ and the scale factor $a_{\rm rad}$ at the radiation time can be readily evaluated using the parametric formulas in the Appendix \ref{app:parametric_formulas}. 

\section{Probing the top-quark mass with gravitational waves} \label{sec:probing_top_mass}

    \begin{figure}[tb]
    \centering
        \includegraphics[width=0.47\textwidth]{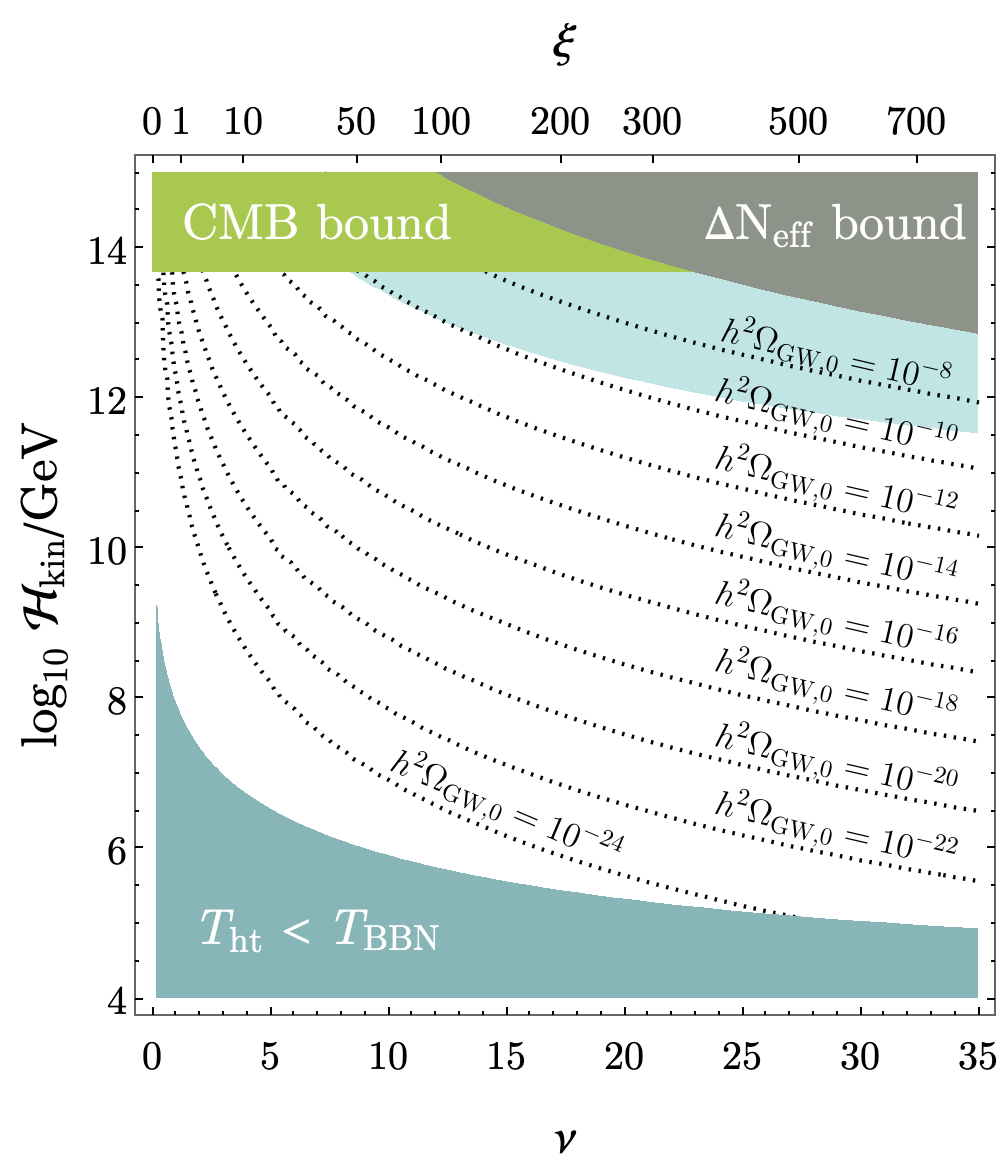}
        \hspace{2mm}
        \includegraphics[width=0.47\textwidth]{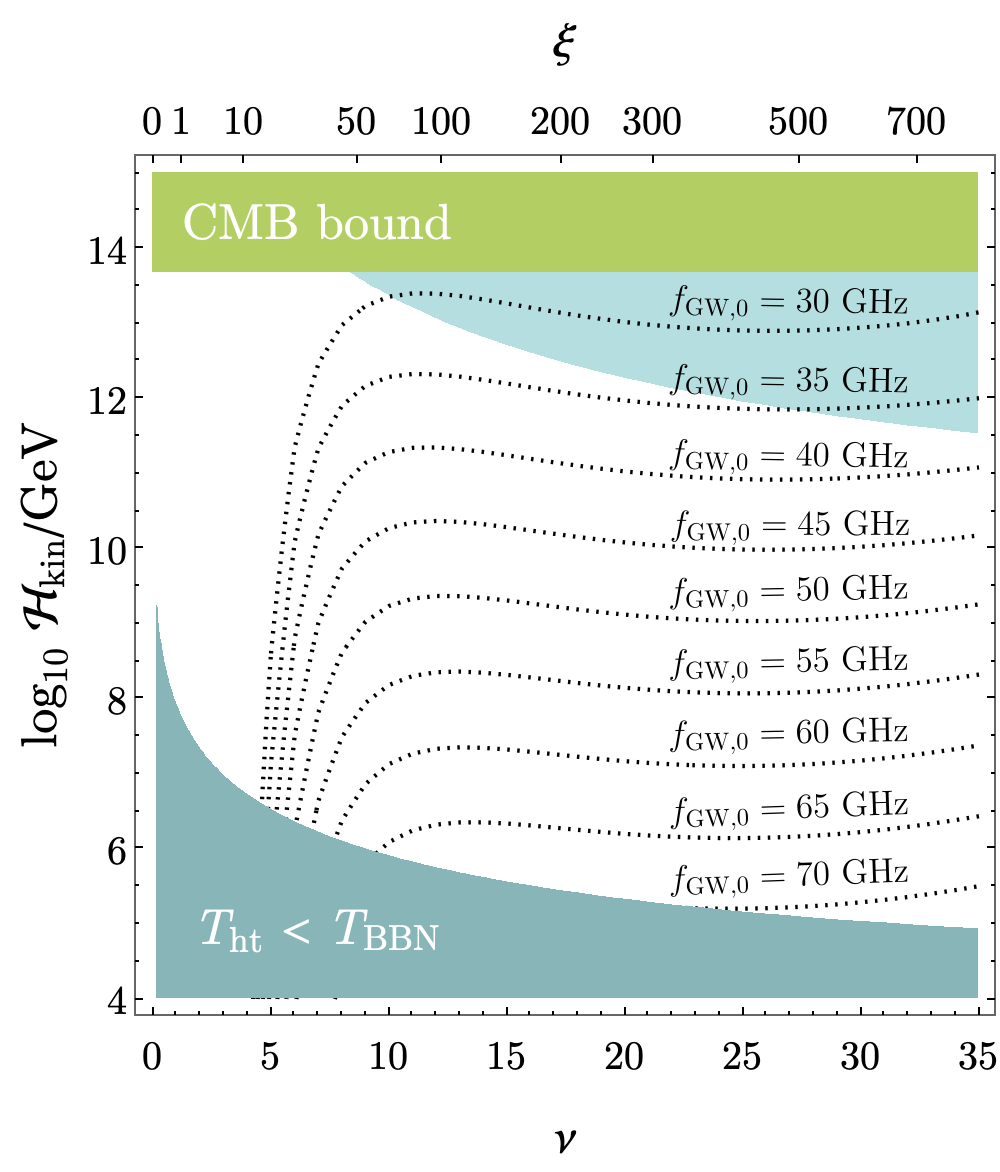}
        \hspace{3mm}
        \caption{Energy density $h^2\Omega_{\rm GW,0}$ and peak frequency $f_{\rm GW,0}$ of the GW spectrum at the present day as a function of $\mathcal{H}_{\rm kin}$ and $\nu$. The top-quark mass has been set to be compatible with the absolute stability of the electroweak vacuum $m_t=170.8 \textrm{ GeV}$. The coloured regions indicate the areas excluded by the proximity to the effective-theory cut-off scale (light blue), by the bound on the inflationary scale from CMB measurements (green), by the number of effective degrees of freedom at BBN (grey), and by the minimum (re)heating temperature (blue).}
        \label{fig:Hkin_GW}
    \end{figure}

In the most minimal scenario, the tachyonic phase in the Higgs dynamics is enough to exponentially enhance the Higgs energy density and lead to the heating of the Universe. The details of the symmetry breaking depend on the running of the Higgs self-coupling in the chosen RGI scheme. By focussing on the dependence on the SM parameter $m_t$ in \eqref{eq:running_fit} we can directly observe the influence that the top-quark mass has on the spectrum of GWs. The specific characteristics of the GW signal enable us to place indirect constraints to the top-quark mass from gravitational physics. 

The existence of an instability scale for larger values of the top-quark mass forces us to consider two distinct possibilities, since the scale of kination has to be a few orders of magnitude smaller than the instability scale to guarantee classical stability via the constraint
\begin{equation}
   \rho_{\text{tac}}(\lambda(h_{\rm min}), \nu) < V_{\rm eff}^{\rm 3 loop}(h_{\rm max}) \; .
\end{equation}
Consequently, if the top-quark mass is large and allows for the negative running of the Higgs self-coupling, the GW amplitude is going to be weaker due to the intrinsically lower kination scale. On the other hand, if the top-quark mass is small and the Higgs is absolutely stable, the phase transition can happen at higher energy scales,~\footnote{
Due to the typical value of ${\lambda(h_{\rm min}) \approx 10^{-3}}$, the parameter space we are considering is generally compatible with the condition $\Delta_1\ll 1$ controlling the actual onset of the HIPT, see the discussion in \cite{Laverda:2023uqv, Laverda:2024qjt}. On the other hand, the condition $\Delta_2\ll 1$ is violated in limited regions of the parameter space where ${\cal H}_{\rm kin}\gtrsim 10^{12}$ and $\nu\gtrsim 10-15$, indicating the breaking of the effective field theory approach. For further details, we refer the reader to Appendix \ref{app:lattice_simulations}.} up to the maximal bound on the inflationary scale $\mathcal{H_{\rm inf}} \leq {\cal H}_{\rm max}=4.7\times10^{13} \textrm{ GeV}$ \cite{Bettoni:2021zhq} set by the Planck Collaboration \cite{Planck:2018jri}, leading to a stronger GW signal. 

\begin{figure}[tb]
\centering
\includegraphics[width=0.85\textwidth]{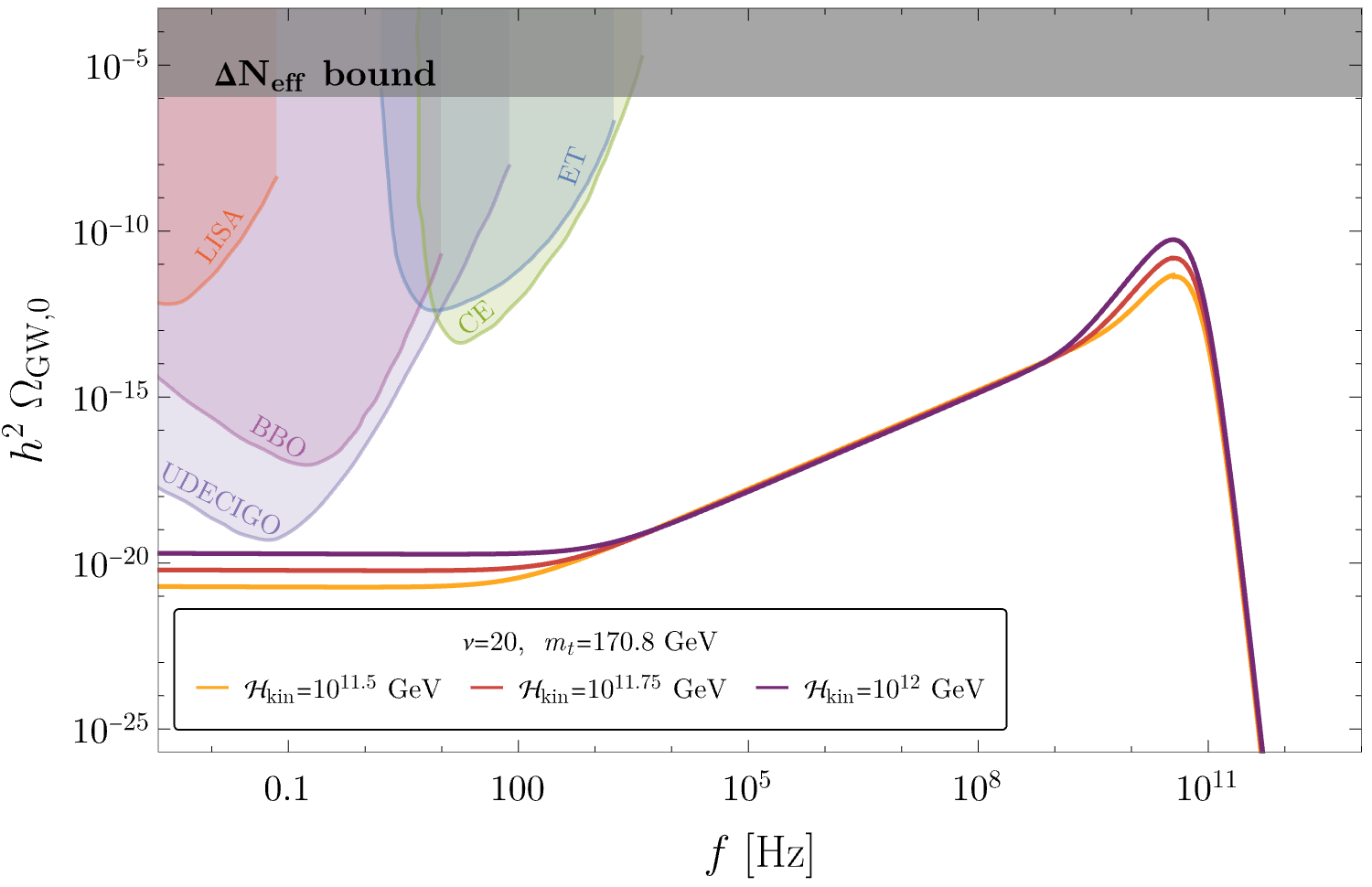}
\hspace{5mm}
\caption{SGWB signal from a Higgs HIPT for different breaking scales compared to sensitivity curves of proposed GW detectors: Laser Interferometer Space Antenna (LISA) \cite{amaroseoane2017laserinterferometerspaceantenna, Robson:2018ifk}, Big Bang Observer (BBO) \cite{Crowder:2005nr, Corbin:2005ny}, UltimateDECIGO \cite{Seto:2001qf, Yagi:2011wg, Kawamura:2020pcg}, Einstein Telescope (ET) \cite{Punturo:2010zz, Branchesi:2023mws}, and Cosmic Explorer (CE) \cite{LIGOScientific:2016wof, Reitze:2019iox}. The non-minimal coupling parameter is set to $\nu=20$ while the top-quark mass $m_t=170.8 \textrm{ GeV}$ ensures the absolute stability of the Higgs vacuum.} \label{fig:GW_absolute stability}
\end{figure}

The threshold separating these two scenarios is easily found from the running in \eqref{eq:running_fit}. The solutions for $\lambda(\mu)=0$ have a bifurcation point at
\begin{equation}
    \Bar{m}_t=170.886 \textrm{ GeV}, \qquad \mu_{\rm inst}=4.13\times 10^{17} \textrm{ GeV}\,,
\end{equation}
giving us the largest top-quark mass $\Bar{m}_t$ that is still compatible with the absolute stability of the SM vacuum. Below this threshold, the kination scale can be a priori as high as $\mathcal{H}_{\rm max}$ but several constraints must be taken into account: first, the self-consistency of the effective field theory approach enforces us to establish an upper cut-off $M_P/\xi$ on the amplitude of the Higgs fluctuations \cite{Barbon:2009ya,Burgess:2009ea,Hertzberg:2010dc,Burgess:2010zq,Bezrukov:2010jz,Bezrukov:2012hx,Ren:2014sya,Bezrukov:2014ipa,Fumagalli:2017cdo,Antoniadis:2021axu,Mikura:2021clt,Ito:2021ssc,Escriva:2016cwl,Karananas:2022byw}; second, the heating of the Universe and the associated onset of radiation domination must be achieved at temperatures above the BBN temperature; third, the contribution of GWs to the total energy density cannot exceed the percentage allowed by the uncertainty on the effective number of relativistic degrees of freedom at BBN \cite{Allahverdi:2020bys,  Caprini:2018mtu, Bettoni:2021zhq},
\begin{equation} \label{eq:Neff_bound}
    h^2 \int \frac{df}{f} \Omega_{\rm GW,0} \lesssim  5.6 \times 10^{-6} \Delta N_{\rm eff} = 1.1 \times 10^{-6}\,.
\end{equation}   
The combined effect of these consistency and viability constraints can be seen in Figure \ref{fig:Hkin_GW}. High symmetry-breaking scales are excluded by the CMB bound and the self-consistency of the scenario, where the light-blue-shaded area excludes the parameter space leading to amplitudes $\langle h^2_{\rm br}\rangle > 10^{-2} \;M_P^2/\xi^2$, while low-scale inflationary models are excluded from heating arguments. Notice that the heating constraint is universal for any value of the top-quark mass, since the Higgs self-coupling converges towards its IR value $\lambda\approx0.1$ at low-energy scales. Dotted lines indicate the energy density of the GW spectrum for different combinations of the parameters $\mathcal{H}_{\rm kin}$ and $\nu$, while the top-quark mass $m_t=170.8 \textrm{ GeV}$ guarantees the absolute stability of the SM vacuum. Interestingly, for $10\lesssim\nu\lesssim30$ the peak frequency of the GW spectrum is almost independent of the non-minimal coupling to gravity and directly correlates a particular value of the top-quark mass to a specific value of the kination scale $\mathcal{H}_{\rm kin}$.

\subsection{Absolutely stable electroweak vacuum} 

\begin{figure}[tb]
    \centering
        \includegraphics[width=0.47\textwidth]{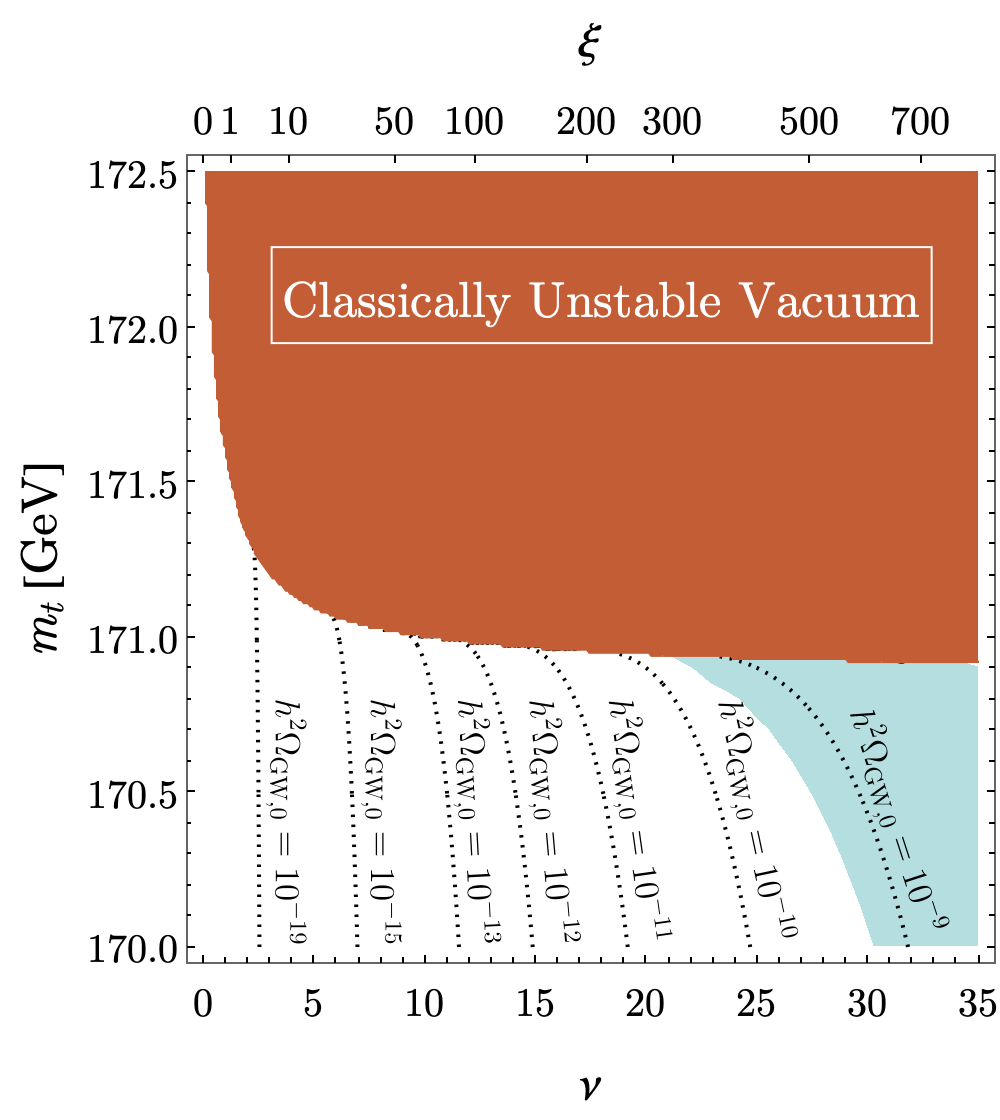}
        \hspace{1mm}
        \includegraphics[width=0.47\textwidth]{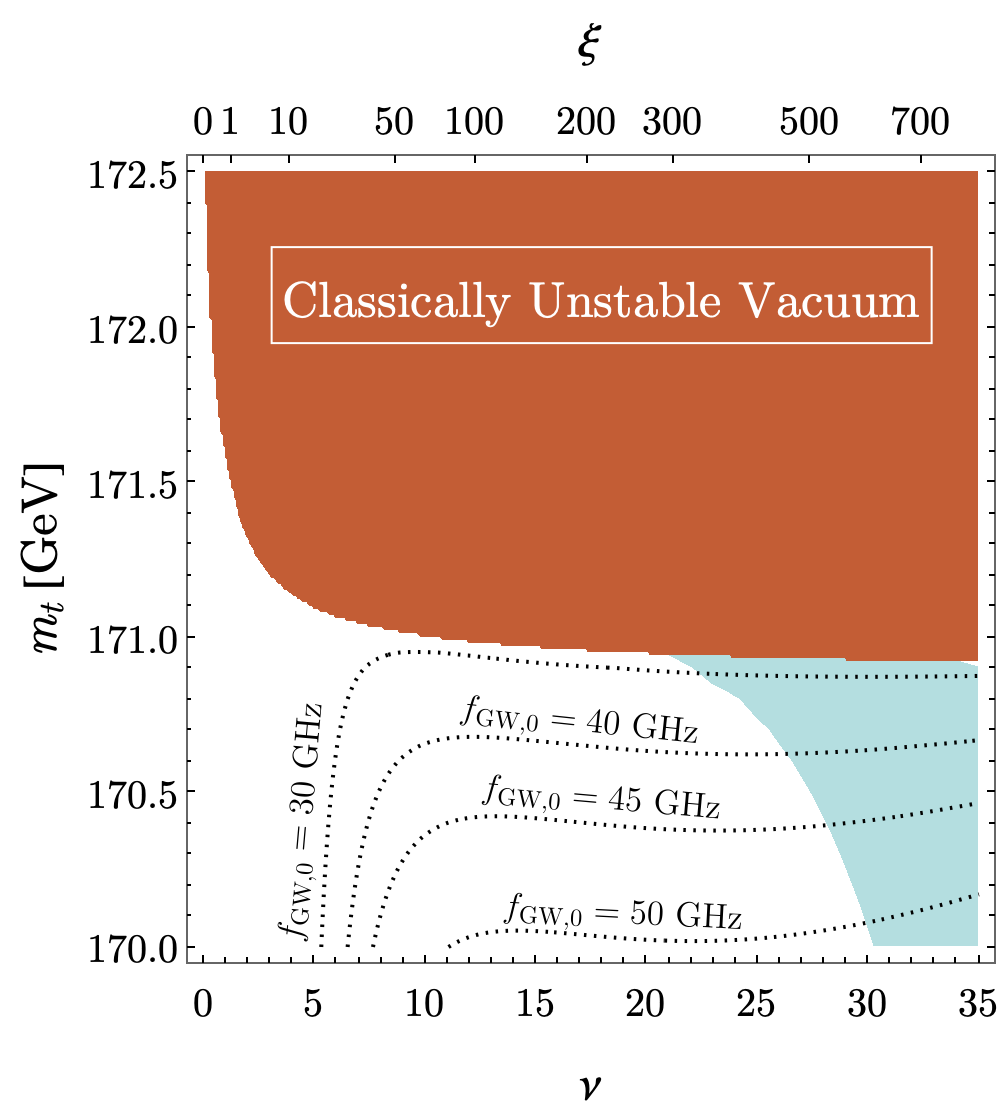}
        \hspace{4mm}
        \caption{Energy density $h^2\Omega_{\rm GW,0}$ and peak frequency $f_{\rm GW,0}$ of the GW spectrum at the present day for different values of the top-quark mass and the non-minimal coupling parameter, with a fixed phase-transition scale $\mathcal{H}_{\rm kin}=10^{12} \text{ GeV}$. Red regions are forbidden because classically unstable, while light-blue shaded areas do not satisfy the 1\% upper bound on the gravitational theory cut-off scale $\langle h^2 \rangle < 10^{-2}  M_P^2 / \xi^2 $.} \label{fig:res_cav_top_mass}
\end{figure}

The full GW signal in the case of absolute stability $m_t<\Bar{m}_t$ is displayed in Figure \ref{fig:GW_absolute stability} for a subset of the parameter space with $\nu=20$. Higher scales of kination produce stronger signals, as it is clear from the fitting formula for the GW amplitude \eqref{eq:fit_momentum_peak} and its rescaling \eqref{eq:f_0}. The typical frequency of the spectra from the phase transition is always within the range $10^{10}-10^{11} \textrm{ Hz}$, restricting the possible detection of the background to ultra-high frequency detectors.\footnote{For a comprehensive analysis on the prospects of high-frequency GW detectors, see e.g.~\cite{Gatti:2024mde, Aggarwal:2020olq, Aggarwal:2025noe}, and \cite{TitoDAgnolo:2024uku} for a recent evaluation of their intrinsic limitations.} On the other hand, the low-frequency part of the spectrum is dominated by the super-horizon inflationary tensor perturbations that are amplified as they reenter the horizon during the period of stiff expansion \cite{Allahverdi:2020bys}. Their contribution to the spectrum is given by
\begin{equation}
    \Omega_{\text{GW, 0}}^{\rm inf}(f) \simeq 10^{-16}\left(\frac{\mathcal{H}_{\rm kin}}{ \mathcal{H}_{\rm max}}\right)^2\left(\frac{f}{f_{\rm pivot}}\right)^{n_t} \; ,
\end{equation}
for modes reentering during radiation-domination, and by  
\begin{equation} 
    \Omega_{\text{GW, 0}}^{\rm inf}(f) \simeq 10^{-16}\left(\frac{\mathcal{H}_{\rm kin}}{ \mathcal{H}_{\rm max}}\right)^2\left(\frac{f}{f_{\rm pivot}}\right)^{n_t} \left({\frac{f}{f_{\rm ht}}}\right) \; ,
\end{equation}
for modes reentering during the stiff phase of kination, with $k_{\rm pivot}=0.002 \text{ Mpc}^{-1}$ the pivot scale at which $\mathcal{H}_{\rm max}$ is measured \cite{Planck:2018jri}, $f_{\rm pivot}$ the frequency of the pivot scale, $f_{\rm ht}$ the frequency corresponding to the Hubble scale at the end of heating and $n_t$ the spectral tilt of inflationary perturbations. It is interesting to notice that the IR stochastic inflationary background could be within reach of low-frequency detectors for sufficiently high kination scales, even though in this case the self-consistency of the procedure becomes potentially compromised. In this regime, two detection windows could simultaneously measure the kination scale in the low frequencies and the characteristics of the spectrum at higher frequencies. In this way, the full parameter space $(\mathcal{H}_{\rm kin}, m_t, \nu)$ would be reduced to $(m_t, \nu)$ with a precise relation between $m_t$ and $\nu$. However, breaking this last degeneracy is a more complicated task that requires either the detection of the peak of the phase-transition signal or the detection of the knees in the spectra indicating the size of the Hubble scale at heating. Indeed, while the peak frequency is mostly dependent on the top-quark mass, the non-minimal coupling parameter $\nu$ is the main actor in the duration of the heating stage, see \cite{Laverda:2024qjt}. The combination of a GW detection with constraints on the (re)heating temperature would, therefore, be able to characterise the Higgs effective potential at high scales.

\subsection{Unstable electroweak vacuum} 
For top-quark masses $m_t>\Bar{m}_t$, enforcing the stability condition in \eqref{eq:barrier_overcoming} sets an upper bound on the kination scale \cite{Laverda:2024qjt} and on the allowed top-mass range, see Figure \ref{fig:res_cav_top_mass}. The instability region is determined by considering only the classical fluctuations of the Higgs field without taking into account the quantum tunnelling between vacua, which could further destabilise the low-scale vacuum. A quick estimate based on the Hawking-Moss and Coleman-de Luccia solutions led to the conclusion that the electroweak vacuum lifetime is primarily affected by the classical overtaking of the barrier in the effective potential \cite{Laverda:2024qjt}. The task of computing a more realistic tunnelling rate in a time-dependent stochastic system with out-of-equilibrium fluctuations is an interesting but complicated problem by itself that goes beyond the scope of the present work \cite{Steingasser:2023gde, Steingasser:2024ikl, Garbrecht:2024end}. 

As can be seen in Figure \ref{fig:res_cav_top_mass}, the GW signal in the case of large top-quark masses is quenched by the small gravitational effective mass. For a fiducial value of ${\mathcal{H}_{\rm kin}=10^{12} \textrm{ GeV}}$ the HIPT spectrum is typically below $h^2\Omega_{\rm GW,0}\sim10^{-9}$ and of comparable magnitude to the SGWB of amplified inflationary perturbations. Once more, it is possible to observe the tight relation between the top-quark mass and the frequency of the GW signal, as noted also in Figure \ref{fig:Hkin_GW}. This feature originates from the dependence of the peak frequency on the rescattering effects taking place in the non-linear phase. Larger values of $\lambda(\mu)$ lead to a quick cascade to smaller scales that produces a GW signal at higher frequencies. In this sense, a direct effect of the running of $\lambda$ and the kination scale can leave an imprint on the GW signal.  
\begin{figure}[tb]
\centering
\includegraphics[width=0.85\textwidth]{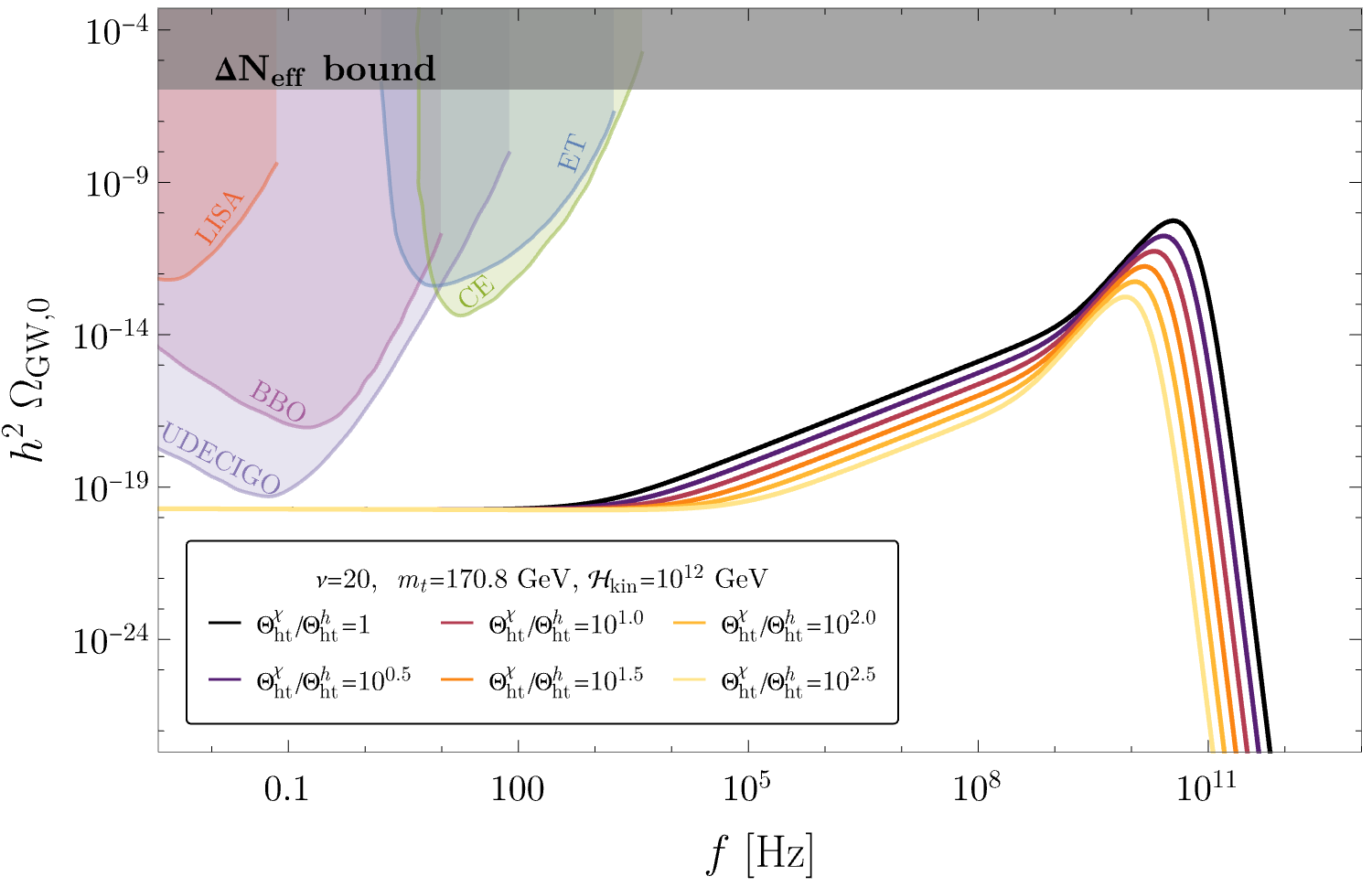}
\hspace{5mm}
\caption{GW signal from the non-minimally coupled Higgs for different heating efficiencies of an enhanced heating sector $\Theta_{\text{ht}}^{\chi}$ compared to sensitivity curves of proposed GW detectors. The model parameters have been set to $\nu=20$ and $m_t=170.8 \textrm{ GeV}$ for a kination scale of $\mathcal{H}_{\rm kin}=10^{12} \textrm{ GeV}$.} 
\label{fig:GW_external_heating}
\end{figure}

Because of the stability bounds enforced on $\mathcal{H}_{\rm kin}$, the frequency range going from kHz to GHz is dominated by the blue-tilted inflationary spectrum, scaling proportionally to $f$ instead of the typical $f^3$-scaling of IR modes excited by causal sources \cite{Caprini:2009fx, Cai:2019cdl}. Indeed, in Figure \ref{fig:res_cav_top_mass} the non-shaded area is allowed by stability constraints but it is characterised by a weaker signal generally obscured by the amplified inflationary perturbations. A precise evaluation of this situation requires a numerical analysis taking into account the horizon-reentering tensor perturbations during inflation, which act as a source term for the GW equation. This goes beyond the numerical simulations performed in \cite{Bettoni:2024ixe}, where a large separation of scales between the kination scale and the typical excited momenta is present. Such condition fails typically for very inefficient tachyonic production at low energy-scales or small non-minimal coupling parameters.

\subsection{Enhanced heating mechanism} \label{sec:probing_reheating}

Although the tachyonic particle production in a HIPT scenario is a very efficient way to achieve a reheated Universe in non-oscillatory models of inflation, it is interesting to consider the effects on the GW signal associated to a different heating mechanism. This is particularly important for ensuring a successful heating at low inflationary scales, $\mathcal{H}_{\rm kin} \lesssim 10^{5.5} \text{ GeV}$, where the radiation temperature of the Higgs sector alone is generically below $T_{\rm BBN}\simeq5 \text{ MeV}$. The specific details of a fully independent heating mechanism are not needed, as long as there exists a radiation-like sector whose energy density is greater than the Higgs'. Even within our framework, two effects can lead to a larger heating efficiency than the one expected from the phase transition alone. In our starting assumptions we have neglected the additional effect that a tree-level coupling to the inflaton field might have on the Higgs fluctuations. For a generic quartic interaction $\sim g^2h^2\phi^2$ the non-adiabatic transition to kination can produce an initial abundance of Higgs particles \cite{Bettoni:2021qfs} thus leading to overall higher Higgs energy density. \footnote{To ensure that the Higgs mass remains independent of the inflaton field in the late Universe, one must assume that, following the transition from inflation to kination, this quartic interaction either becomes negligible or is forbidden by symmetry considerations \cite{Bettoni:2021qfs, Rubio:2017gty}.} On top of this effect, the non-instantaneous change in the Ricci scalar can also lead to some transient amplification of the Higgs fluctuations, a factor that is more important for small values of $\xi$.

Regardless of its fundamental nature, the heating efficiency language allows us to parametrise the macroscopic characteristics of an enhanced reheating sector, which we label with $\chi$, whose heating efficiency is simply $\Theta_{\text{ht}}^{\chi} \equiv \rho_{\chi}(a_{\rm rad})/ \rho_{\phi}(a_{\rm rad})$. This quantity is the only additional free parameter we need to consider in our analysis. Since $\Theta_{\text{ht}}^{\chi} \geq \Theta_{\text{ht}}^{h}$ by construction, the (re)heating temperature for an enhanced heating is higher than the Higgs' HIPT scenario,
\begin{equation}
T_{\rm ht}^{\chi} =\left(\frac{30\,\rho^{\rm ht}_{\chi}}{\pi^2 g_*^{\rm ht}}\right)^{1/4} = \left( \frac{\Theta_{\text{ht}}^{\chi}}{\Theta_{\text{ht}}^{ h}} \right)^{1/4} \times T_{\rm ht}^{ h} \,,   
\end{equation}
and kination is shorter-lived than in the standard case. The scaling of the GW
signal from the moment of production to the present day is also affected, with a redshift in frequencies and a decrease in amplitude, as one can easily see in the formulas 
\begin{gather}
    f_{\rm GW,0} (\nu, m_t, \Theta_{\text{ht}}^{\chi})\simeq 1.3\times 10^9 \, \text{Hz} \; \frac{\kappa}{2\pi} \left(\frac{\mathcal{H}_{\rm kin} \, a_{\rm rad}}{10^{10} \textrm{ GeV}}\right)^{1/2} \left( \frac{\Theta_{\rm ht}^{\chi}}{10^{-8}}\right)^{-1/4} \,, \nonumber \\
    \Omega_{\rm GW,0} (\mathcal{H}_{\rm kin},\nu, m_t, \Theta_{\text{ht}}^{\chi}) = 1.67 \times 10^{-5} h^{-2} \left( \frac{100}{g_*^{\rm ht}} \right)^{1/3} \left( \frac{\Theta_{\text{ht}}^{h}}{\Theta_{\text{ht}}^{\chi}} \right) \times \Bar{\Omega}_{\rm GW}\,.
\end{gather}
Notice that we recover the Higgs HIPT result if we assume $\Theta_{\rm ht}^h=\Theta_{\rm ht}^\chi$. In Figure \ref{fig:GW_external_heating}, the combined effects of an augmented heating sector are visible in the relative positioning of the GW spectra. All model parameters have been fixed except from $\Theta_{\rm ht}^\chi$, which can take on values much larger than $\Theta_{\rm ht}^h$. For highly efficient heating, the spectrum is peaked at lower frequencies but at the same time its amplitude is reduced, since the HIPT Higgs is energetically subdominant with respect to the more efficient heating scenario. The $\chi$-sector breaks the scaling relations that characterise the Higgs HIPT scenario by introducing a new independent energy scale. The potential detection of a peaked GW signal at frequencies below the typical range in Figure \ref{fig:GW_absolute stability} could be a smoking-gun clue for additional new physics related to direct inflaton couplings or external heating sectors beyond the SM. 

\section{Stability beyond the Standard Model} \label{sec:bsm_stability}

As shown in Appendix \ref{app:cut-off}, the non-minimal coupling of the Higgs field to gravity makes the model under consideration intrinsically non-renormalisable, requiring it to be treated as an effective field theory valid up to a specific cut-off scale, $\Lambda_\xi = M_P /\xi$ \cite{Barbon:2009ya,Burgess:2009ea,Hertzberg:2010dc,Burgess:2010zq,Bezrukov:2010jz,Bezrukov:2012hx,Ren:2014sya,Bezrukov:2014ipa,Fumagalli:2017cdo,Antoniadis:2021axu,Mikura:2021clt,Ito:2021ssc,Escriva:2016cwl,Karananas:2022byw}, which may either signal the transition to a strongly coupled regime \cite{Aydemir:2012nz,Calmet:2013hia,Escriva:2016cwl,Saltas:2015vsc} or the emergence of new degrees of freedom beyond the SM content \cite{Giudice:2010ka} (for a review in the context of Higgs inflation see \cite{Rubio:2018ogq}).  For the non-minimal couplings considered in this work, $\Lambda_\xi$ is estimated to range from approximately between $10^{16}$ to $10^{17}$ GeV, i.e. typically well above the SM vacuum instability scale. However, the validity of the low-energy effective field theory could be further reduced in the presence of additional sub-Planckian degrees of freedom beyond the SM content.

\begin{figure}[tb]
\centering
\includegraphics[width=0.6\textwidth]{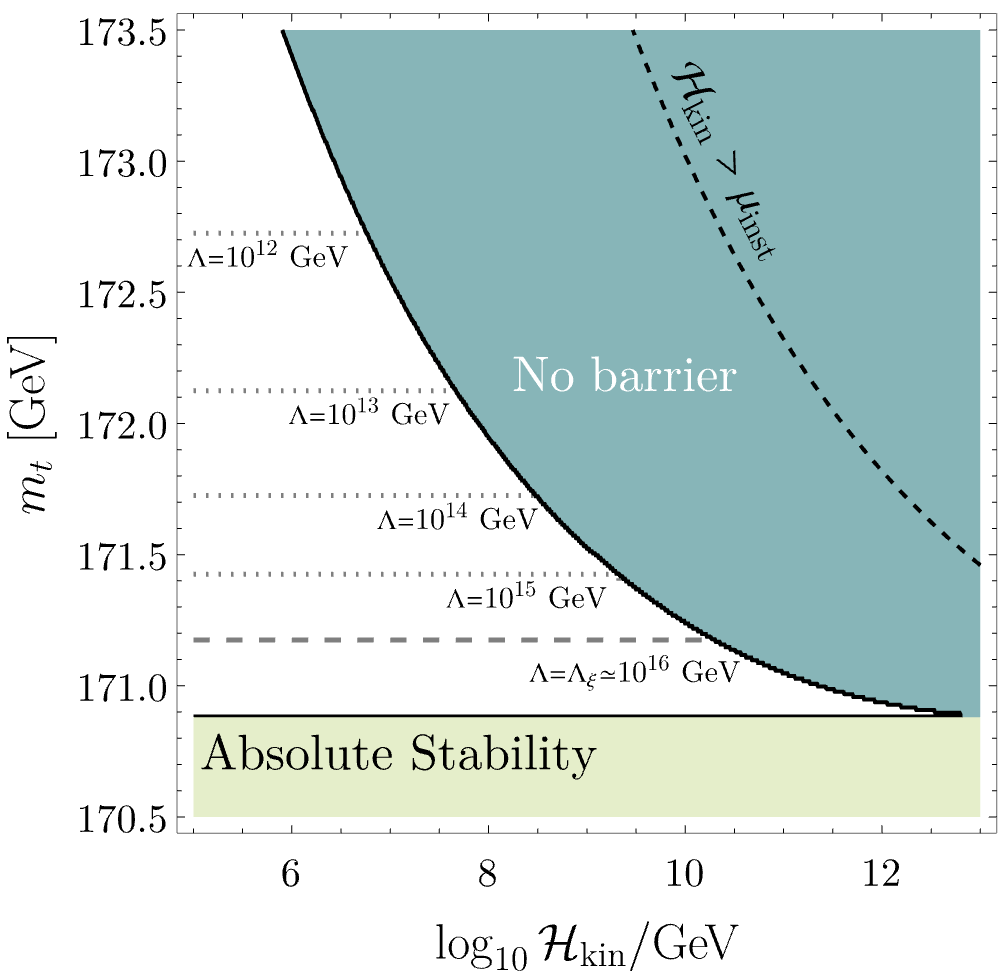}
\caption{Impact of non-renormalisable sextic operators in the vacuum structure of the Higgs effective potential. The blue area indicates the parameter space where the non-minimal curvature interaction suppresses the quartic barrier. The light-green area indicates absolute stability for $m_t\leq\bar{m}_t$, where only the low-scale minimum is present. The metastable scenario involving two minima is realised in the white area for the general case of the Standard-Model only effective potential. Including the sextic interaction term $(H^{\dagger}H)^3/\Lambda^2$ leads to a shift in the absolute stability bound, indicated by the dotted lines for different values of the cut-off scale $\Lambda$. The model parameters have been fixed to $m_h=125.2 \textrm{ GeV}$ and $\nu=20$, leading to an effective-field-theory cut-off scale $\Lambda_{\xi}=9.2\times10^{15} \textrm{ GeV}$ (dashed line).} \label{fig:mt_Hkin_instability_scale}
\end{figure}

From a bottom-up perspective, such new physics would manifest itself as higher-dimensional operators suppressed by a cut-off scale $\Lambda \lesssim \Lambda_\xi$, determined by the thresholds and masses of the new particles that are integrated out during the construction of the low-energy effective field theory. Interestingly enough, the inclusion of these so-far omitted non-renormalisable operators could potentially restore the convexity of the Higgs effective potential at high energies (see Figure \ref{fig:effective_potential}), leading to the absolute stability of the electroweak vacuum irrespectively of the specific top-quark contribution. To investigate this possibility, we consider the impact of a prototypical non-renormalisable sextic operator, constructed from the gauge-invariant combination $H^\dagger H$ and the hypothetical cut-off scale of new physics $\Lambda$ as free parameter. The corresponding effective potential takes the form
\begin{equation} 
    V_{\rm eff}(H) = \xi R H^{\dagger}H + \lambda(H^{\dagger}H)^2 + \frac{1}{\Lambda^2} (H^{\dagger}H)^3\;.
    \label{eq:effective_potential_sextic}
\end{equation}
Note that the inclusion of the sextic operator is expected to modify the $\beta$-functions of the theory \cite{Eichhorn:2015kea}, requiring their rederivation for a fully rigorous treatment of the problem. However, since the threshold effects associated with non-renormalisable operators become significant only in the vicinity of the characteristic scale, we will proceed by retaining the original three-loop running in \eqref{eq:running_fit} while incorporating the sextic operator as a simple tree-level term, thereby neglecting its energy dependence \cite{Branchina:2013jra}. 

\begin{figure}[tb]
\centering
\includegraphics[width=0.65\textwidth]{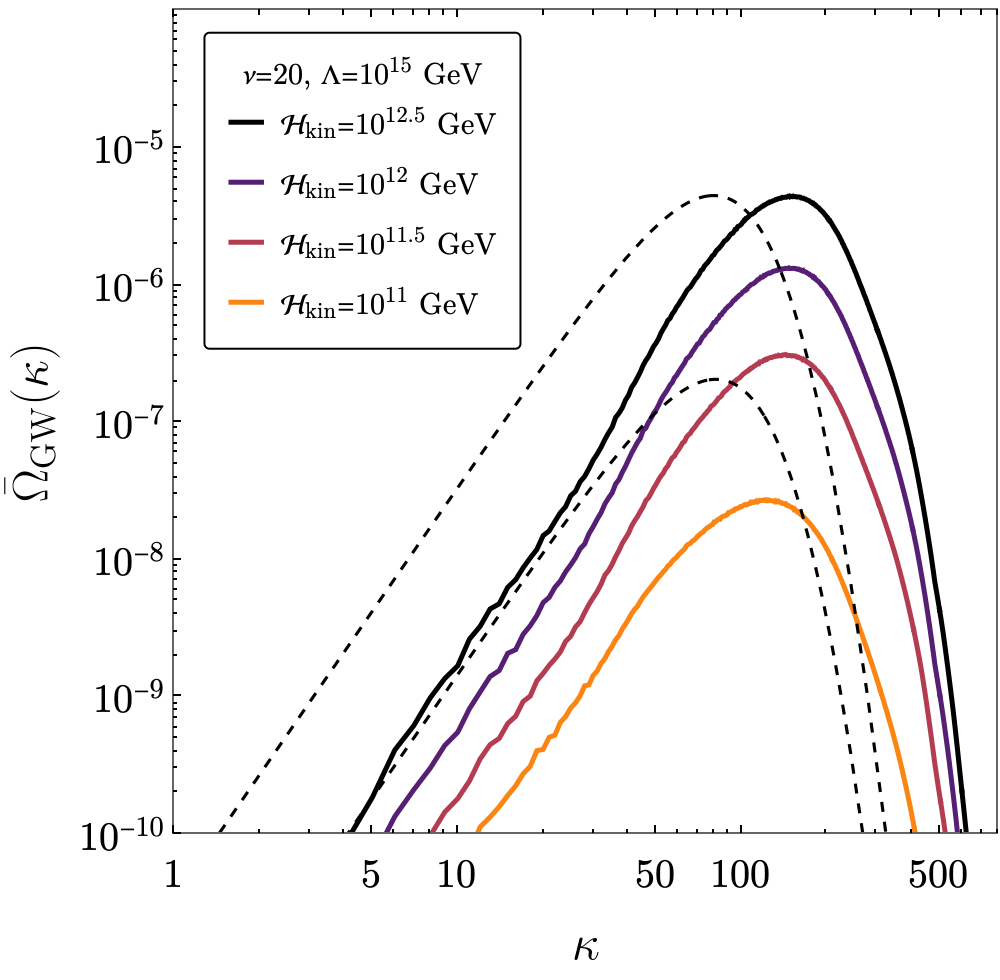}
\caption{GW signal after the end of the production phase in the scenario involving a non-minimally coupled Higgs with sextic self-interactions in the effective potential. Different colours correspond to different choices of the kination scale $\mathcal{H}_{\rm kin}$ in our benchmark simulations, while the cut-off scale of new physics has been set to $\Lambda=10^{15} \textrm{ GeV}$. The remaining model parameters have been set to $\nu=20$ and $m_t=171.3 \textrm{ GeV}$. Dashed lines indicate the maximum and minimum GW signal shown in Figure \ref{fig:GW_absolute stability} with $m_t=170.8 \textrm{ GeV}$ and $\mathcal{H}_{\rm kin}=10^{11.5}, \, 10^{12} \textrm{ GeV} $. Momenta are normalised with respect to $\mathcal{H}_{\rm kin}$ and the simulations are well within the effective-field theory cut-off, namely $\langle h^2 \rangle <  10^{-2} M_P^2/\xi^2 $.} \label{fig:gw_sextic}
\end{figure}

As illustrated in Figure \ref{fig:mt_Hkin_instability_scale}, the presence of new physics at the scale $\Lambda$ shifts the super-Planckian minimum of the effective potential to smaller scales, or even eliminates it entirely. This figure provides a direct representation of the different phases of the model at the time of symmetry breaking, i.e. at the onset of kinetic domination. In the absence of new physics, the classical dynamics of the Higgs fields must be generally checked for instability within the parameter space corresponding to the white area, while absolute stability is ensured only for top-quark masses below the critical value of $\Bar m_t=170.886 \textrm{ GeV}$. If the inflationary scale $\mathcal{H}_{\rm kin}$ is sufficiently close to the instability scale $\mu_{\rm inst}$, the potential lacks a barrier due to the large contribution from the non-minimal coupling and only one high-scale vacuum is present. The inclusion of sextic interactions stabilises the potential by restoring convexity at sub-Planckian scales. When a barrier is present, a large cut-off scale $\Lambda$ is enough to ensure stability for low top-quark masses, while a smaller $\Lambda$ can stabilise the potential for a wider range of masses. As a result, the critical boundary at $\Bar m_t$ shifts upwards, as indicated by the dashed lines in the figure, thus expanding the parameter space associated with absolute stability. When the barrier is fully suppressed, the higher-order operator provides the presence of a high-scale vacuum below the Planck mass. Notice that this analysis concerns only the overall vacuum structure of the potential at the time of symmetry breaking for a specific value of non-minimal coupling parameter and does not include the Higgs dynamics.\footnote{In general, for the parameter space under consideration, the Hubble scale at the end of the first semi-oscillation of the field is about one order of magnitude smaller than $\mathcal{H}_{\rm kin}$. Qualitatively speaking, this difference shifts the black boundary in Figure \ref{fig:mt_Hkin_instability_scale} to higher values of $\mathcal{H}_{\rm kin}$.} 

Within the newly stabilised region, we turn once again to 3+1 classical lattice simulations. The main advantage of this numerical technique consists in the accurate description of the non-linear field dynamics and the prediction of the associated GW signal. The setup behind the implementation via the code \texttt{$\mathcal{C}osmo\mathcal{L}attice$} \cite{Figueroa:2020rrl, Figueroa:2021yhd} is described in Appendix \ref{app:lattice_simulations}. For this discussion, it suffices to state that the numerical model incorporates the complete potential in \eqref{eq:effective_potential_sextic}, along with its parametric three-loop running \eqref{eq:running_fit}. The simulations are initialised at the time of symmetry breaking, i.e. at the transition between inflation and kination, and evolved for at least two e-folds on a kination background to capture the GW spectrum after the end of the non-linear regime. This implementation follows the established methods already tested in previous studies, including the SM Higgs scenario without new physics \cite{Laverda:2024qjt} and the prototypical case of a scalar singlet at tree level \cite{Laverda:2023uqv, Bettoni:2024ixe}. We consider four benchmark points for $\mathcal{H}_{\rm kin} \in [10^{11}, \, 10^{12.5} \textrm{ GeV}]$, while we keep the scale of new physics at $\Lambda=10^{15} \textrm{ GeV}$. In order to assure the covering of all relevant momenta, we perform simulations on a large grid of $N=960$ lattice points per side.

The GW signal generated by the phase transition is presented in Figure \ref{fig:gw_sextic}. The spectra are measured after the end of the large-oscillation non-linear phase that follows the tachyonic instability. Indeed, as was seen in \cite{Bettoni:2024ixe}, the GW production is concentrated in this period, since the gradient energy density is at its peak. For IR momenta, we recover the causal spectral slope proportional to $f^3$. However, the frequency range of metric perturbations is shifted to higher values with respect to the GW signal shown in Figure \ref{fig:GW_absolute stability}. Interestingly, the position of the peak evolves from $\kappa\sim 120$ for $\mathcal{H}_{\rm kin}=10^{11} \textrm{ GeV}$ to $\kappa\sim 150$ for $\mathcal{H}_{\rm kin}=10^{12.5} \textrm{ GeV}$. This characteristic evolution with increasing inflationary scales is in contrast with the one expected in the pure SM scenario, where higher inflationary scales lead generally to smaller self-couplings and therefore lower peak frequencies. The difference is explained by the more efficient turbulent cascade \cite{Micha:2004bv} towards UV modes induced by the sextic interaction, which quickly breaks down the large-amplitude fluctuations and in so doing it excites high-frequency modes more intensely than in the SM scenario. Moreover, one can also notice different slopes around the peak, which can constitute a telltale sign of the presence of new physics \cite{Dufaux:2010cf}. Indeed, for high inflationary scales, the sextic operator becomes essential in guaranteeing the stability of the SM Higgs field and the stochastic system reflects the presence of such additional interactions. These bespoke features can shed light on the presence of non-renormalisable operators in the Higgs potential via a possible GW detection at very high frequencies. 

\section{Conclusions} \label{sec:conclusion}

Vacuum stability is one of the most profound questions in fundamental physics, directly tied to the structure of the Higgs potential and the long-term fate of our Universe. One of the most significant outcomes of the Large Hadron Collider has been its ability to probe the stability of the electroweak vacuum by precisely measuring the Higgs boson and top-quark masses, two key parameters that determine whether the vacuum is absolutely stable, metastable, or prone to decay over cosmological timescales. Intriguingly, current measurements place the SM vacuum near a critical boundary between stability and metastability, suggesting the possibility of new physics at higher energy scales. This delicate balance opens a compelling avenue for exploring BSM physics, as even slight modifications to the Higgs potential, such as new scalar fields, non-renormalisable interactions, or quantum gravity effects, could dramatically alter the fate of the vacuum.

The present work aims at probing the high-scale Higgs effective potential via GWs. The interplay of non-minimal gravitational couplings and a kination epoch results in a Hubble-induced phase transition that quickly reheats the Universe in a non-perturbative way while leaving a specific top-mass-dependent GW signature. We have shown that top-quark masses compatible with absolute stability generate a strong signal, while higher top-quark masses enforce more stringent upper bounds on the inflationary scale, thus weakening the overall signal. A simultaneous detection of the inflationary background at low-frequencies and of the peak of the phase transition's spectrum at high frequencies can quantify the model parameters. In this sense, the scenario under consideration offers a way to investigate gravitationally the magnitude of the top-quark mass. Notably, while the peak frequency of the spectrum is strongly correlated with the top-quark mass, the (re)heating temperature is predominantly determined by the effective tachyonic mass. Different cosmological constraints can act complementarily in characterising the gravitational and Standard-Model couplings of the Higgs at high scales.  

A key aspect of the analysis is the extensive use of parametric formulas, derived from hundreds of 3+1 classical lattice simulations, which enable us to predict the general shape of the SGWB as a function of the top-quark mass, being this quantity the largest source of uncertainty in the stability of the electroweak vacuum. It should be noted that, unlike other works on GWs from vacuum instability \cite{Espinosa:2018eve}, our setup does not rely on the temporary instability of the Higgs vacuum to generate a GW signal. On the contrary, we have shown that a GW signal is unavoidable even in the case of absolute stability because of the combination of non-minimal curvature couplings and a stiff post-inflationary expansion phase.

We have also speculated about the possibility of BSM physics affecting the Higgs potential through non-renormalisable operators at scales below the cut-off dictated by the self-consistency of the theory. In this case, the positioning of the spectrum and its intrinsic characteristics can allow a direct link between the produced GW signal and the nature of such new interactions. This result can prompt some very interesting questions within the scenario of a non-minimally coupled Higgs field: Could specific patterns in the GW signal serve as a smoking gun for specific BSM extensions, such as Higgs portal models or new scalar fields coupled to the Higgs? Do these new interactions leave unique fingerprints in the Higgs potential, altering its phase structure in ways we can detect? How can we implement a consistent picture of quantum tunnelling in a stochastic system with a non-homogeneous and time-dependent fluctuating Higgs field? Can the interactions between the Higgs field and other thermal bath components leave detectable imprints on the GW spectrum?  We plan to explore these questions in future publications.     

\section*{Acknowledgments}

We extend our gratitude to Dario Bettoni for engaging discussions that contributed to the preparation of this work. We also thank Sebastian Ellis for the valuable comments on the first version of this manuscript. The numerical 3+1 classical lattice simulations were carried out with the support of the Infraestrutura Nacional de Computa\c c\~ao Distribu\'ida (INCD) funded by the Funda\c c\~ao para a Ci\^encia e a Tecnologia (FCT) and FEDER under the project 01/SAICT/2016 nº 022153. G.~L. acknowledges support from a fellowship provided by ``la Caixa” Foundation (ID 100010434) with fellowship code LCF/BQ/DI21/11860024, as well as the support of FCT through the grant with Ref.~2024.05847.BD. G.~L. acknowledges also the financial support from FCT to the Center for Astrophysics and Gravitation-CENTRA, Instituto Superior Técnico, Universidade de Lisboa, through Project No.~UIDB/00099/2020. J.~R. is supported by a Ram\'on y Cajal contract of the Spanish Ministry of Science and Innovation with Ref.~RYC2020-028870-I. This research was further supported by the project PID2022-139841NB-I00 of MICIU/AEI/10.13039/501100011033 and FEDER, UE. Project carried out with the 2025 Leonardo Grant for Scientific Research and Cultural Creation from the BBVA Foundation. The BBVA Foundation is not responsible for the opinions, comments, and content included in the project and/or its resulting outcomes, which are the sole and exclusive responsibility of the authors

\appendix

\section{Einstein-frame formulation and the EFT cut-off} \label{app:cut-off}

The value of the cut-off scale entering the effective potential \eqref{eq:effective_potential_sextic} is a priori unknown and depends on the various thresholds and masses of potential BSM particles that were integrated out to obtain the low-energy effective field theory. The non-renormalisability of the SM non-minimally coupled to gravity sets, however, an upper bound on this scale for a given value of $\xi$ \cite{Barbon:2009ya,Burgess:2009ea,Hertzberg:2010dc,Burgess:2010zq,Bezrukov:2010jz,Bezrukov:2012hx,Ren:2014sya,Bezrukov:2014ipa,Fumagalli:2017cdo,Antoniadis:2021axu,Mikura:2021clt,Ito:2021ssc,Escriva:2016cwl,Karananas:2022byw}. To see this explicitly, let us consider the action for the inflaton field $\phi$ and the subdominant Higgs field in the unitary gauge $H=(0,\,h/\sqrt{2})^T$, 
\begin{equation} \label{eq:higgs_lagrangian2}
S= \int d^4x \sqrt{-g} \left[\left( \frac{M^2_{P}}{2} - \frac12 \xi h^2 \right) R - \frac12 \partial^\mu h \partial_\mu h- \frac{\lambda}{4} h^4  + \mathcal{L}_{\rm \phi}\right] \,.
\end{equation}
For the sake of concreteness and without any lack of generality, we will assume the inflaton field to display a standard Lagrangian density 
\begin{equation} \label{Linflaton}
\mathcal{L}_{\phi} = - \frac12 \partial^\mu \phi \partial_\mu \phi - V(\phi)\,, 
\end{equation}
with $V(\phi)$ a \textit{runaway or quintessential} potential able to support a sustained-enough period of kinetic domination immediately after the end of inflation. Performing a Weyl rescaling of the metric $g_{\mu\nu}\to \Omega^{-2} g_{\mu\nu}$ with conformal factor $\Omega^2=1-{\xi}h^2/M^2_{P}$, we obtain the equivalent Einstein-frame action
\begin{equation}
S= \int d^4x \sqrt{-g} \left[\frac{M_{P}^2}{2}{R} 
-\frac12 \left( \frac{1}{\Omega^{2}}+\frac{6\xi^2h^2}{M_{P}^2\Omega^{4}} \right)\partial^\mu h \partial_\mu h- \frac{\lambda}{4}\frac{h^4}{\Omega^{4}}
+\frac{1}{\Omega^4}{\tilde{\cal L}}_\phi\right]\,,
\end{equation}
with 
\begin{equation}
 \tilde{\cal L}_{\phi} =-\frac{1}{2}\,\Omega^2 \partial^\mu \phi\partial_\mu \phi-V(\phi)\,.
\end{equation}  
For non-minimal couplings $\xi\gg 1$ and field values $u\equiv \sqrt{\xi} h/M_{P}\ll 1$, this non-linear expression can be written 
\begin{align}
S\simeq \int d^4x \sqrt{-g} \Bigg[  \frac{M_{P}^2}{2} R &+\mathcal{L}_{\phi} - \frac{1}{2}\left(1+ \frac{6h^2}{\Lambda_\xi^2}  \right)\partial^\mu h \partial_\mu h  - \frac{\lambda}{4}h^4  \nonumber \\ 
 & +\frac{1}{2}\frac{\xi h^2}{M^2_{P}} \,T^{(\phi)} +{\cal O}(u^2\,{\cal L}_h, u^4{\cal L}_\phi )
\Bigg] \,,
\label{Eframeaction}
\end{align} 
where we have identified the cut-off-scale of the theory from the leading-order correction ($n=1$) to the Higgs kinetic term at large $\xi$ values,
\begin{equation}
\Lambda_\xi=\frac{M_P}{\xi}\,,
\end{equation}
explicitly isolated the $n=1$ correction to the inflaton Lagrangian density ${\cal L}_\phi$ and introduced the energy-momentum tensor of the inflaton field and its trace,
\begin{equation}
T^{(\phi)}_{\mu\nu} =\partial_\mu\phi\partial_\nu\phi-g_{\mu\nu}\left(\frac12\partial^\rho\phi\partial_\rho\phi+V(\phi)\right)\,,\quad \quad  T^{(\phi)}\equiv T^{(\phi)\,\mu}{}_{\mu} =-\left(\partial^\mu \phi  \partial_\mu \phi+4V(\phi)\right) \,.
\end{equation}
Notice that a hierarchy 
\begin{equation}
    T^{(\phi)} \gg\frac{\xi h^2}{M^2_{P}} \,T^{(\phi)} \gg \frac{\xi h^2}{M^2_{P}} \,\mathcal{L}^{(h)} , \frac{\xi^2 h^4}{M^4_{P}}\mathcal{L}^{(\phi)}
\end{equation}
is realised in the perturbative expansion, which inherently suppresses any direct coupling between the Higgs and the inflaton that might arise in the Einstein frame even when the same tree-level coupling is absent from the Jordan-frame Lagrangian. Although the fifth term on the right-hand side of \eqref{Eframeaction} is divided by $M_P^2$, it is not suppressed relative to the Higgs Lagrangian density for $T^{(\phi)}\gg T_{\xi=0}^{(h)}$. For an homogeneous inflaton field $\phi(t)$ in a flat FLRW background, $T^{(\phi)}$ becomes
\begin{equation}
T^{(\phi)}=-\rho_\phi+3p_\phi=-(1-3 w_\phi)\rho_\phi\,, 
\end{equation}
with $\rho_\phi$ and $p_\phi$ the energy density and pressure of the inflaton field in this regime and $w_\phi=p_\phi/\rho_\phi$ the corresponding equation-of-state parameter. The dynamics associated with the action \eqref{Eframeaction} coincides therefore with that assumed in the main text, provided the effect of the higher-dimensional operator suppressed by the cut-off scale $\Lambda_\xi$ can be neglected. Restricting ourselves to field values $h \ll \Lambda_\xi$, we have
\begin{equation}
    S\simeq \int d^4x \sqrt{-g} \left[ \frac{M^2_P}{2}R + \mathcal{L}_{\phi}  -\frac12{\partial_{\mu}}{h} \partial^{\mu} h -\frac12 \xi R h^2 - \frac{\lambda}{4}h^4 \right] \,, 
\end{equation}
with $R\simeq -T^{\phi}/M_P^2\simeq 3(1-3w_\phi){\cal H}^2$.

\section{Summary of parametric formulas from lattice simulations} \label{app:parametric_formulas}

In this Appendix, we summarise the most relevant quantities that characterise the course of a Hubble-induced phase transition in the Higgs sector. In order to maintain the approach followed in previous works, we consider the tree-level Higgs potential discussed in \cite{Laverda:2023uqv}, with fixed $\lambda$ and $\nu$. There exist two essential timescales in the evolution of the spectator field after symmetry breaking
\begin{itemize}
   \item The backreaction time $z_{\rm br}$ is defined as the time at which self-interaction effects from the quartic potential term become non-negligible and the initial tachyonic phase is brought to an end. We use the following implicit definition
    \begin{equation}
        (-\nabla Y + \lambda Y^3)\eval_{\rm br} = M^2(z)Y \eval_{\rm br}
    \end{equation}
    where the auxiliary field $Y=h/a\sqrt{6\xi}\mathcal{H}_{\rm kin}$ is a rescaled conformal version of the Higgs field in unitary gauge and $M^2(z) = (\nu^2 - 1/4)/(z+\nu)^2$. The previous identity quantifies the time at which the effective frequency of the Higgs fluctuations vanish, thereby coinciding with the condition $Y''(z_{\rm br})=0$. The definition of backreaction time can be easily estimated thanks to the linear analysis performed in \cite{Bettoni:2019dcw}. Here we only quote the main result following the standard procedure of solving mode-by-mode the equation of motion, namely the expression for the evolution of the Higgs fluctuations
    \begin{equation}\label{eq:y2}
        Y^2_{\rm rms}(z)= \left(\frac{2 \nu-1}{8\pi \nu}\right)^2    \left(1+\frac{z}{\nu}\right)^{2 \nu +1} \kappa_*^2(z)\,,
    \end{equation}
    with $\kappa_*=k/\sqrt{6\xi}\mathcal{H}_{\rm kin}=2\sqrt{\nu+1}\, /(2(z+\nu))$ a typical momentum scale. Thanks to a large number of numerical 3+1 classical lattice simulations, it has been possible to characterise the amplitude and energy density of the Higgs field at the backreaction timescale in terms of parametric formulas:
    \begin{gather}
        \langle h^2_{\rm tac} (\lambda, \nu)\rangle = 4 \mathcal{H}^2_{\rm kin} \, \exp\left(\alpha_1 + \alpha_2 \nu +{\alpha_3} \log \nu \right)\, ,\\
        \alpha_1 = -4.92 + 0.74 \, n  \,, \;  \alpha_2 = -0.04 - 0.02 \, n \,, \; \alpha_3 = 3.54 + 0.61 \, n \,,
    \end{gather}
    and 
    \begin{gather}
        \rho_{\text{tac}}(\lambda, \nu) = 16 \, \mathcal{H}^4_{\rm kin}\, \exp\left(\beta_1 + \beta_2 \,\nu+ {\beta_3} \log \nu \right)  \,,\\
        \beta_1 = -7.03 - 0.56 \, n  \,, \;
        \beta_2 = -0.06 - 0.04 \, n \,, \;
        \beta_3 = 7.15 + 1.10 \, n \,, \;
    \end{gather} 
    with $\nu=\sqrt{3\xi/2}$ and $n=-\log_{10}(\lambda)$.
    \item In the same lattice-based analysis, it is possible to derive the time at which the Higgs energy density achieves a radiation-like scaling $z_{\rm rad}$ and evaluate its energy density at that point. We found
    \begin{gather}
        z_{\text{rad}}(\lambda, \nu) = \gamma_1 + \gamma_2 \, \nu \,, \\
        \gamma_1 = 33.63 + 15.02 \, n - 0.22 \, n^2 \,, \;
        \gamma_2 = 7.91 - 0.01 \, n + 0.02 \, n^2 \,. \;
    \end{gather}   
    The Higgs energy density at that timescale is given by
    \begin{gather}
        \rho_{\text{rad}}(\lambda, \nu) = 16 \mathcal{H}^4_{\rm kin} \, \exp\left({\delta_1 + \delta_2 \, \nu} +{\delta_3}\log \nu\right) \,, \\
        \delta_1 = -11.10 - 0.06 \, n \,, \;
        \delta_2 = -0.04 - 0.03 \, n \,,  \;
        \delta_3 = 5.62 + 0.87 \, n \,. \;
    \end{gather}  
    Note that $\mathds{Z}_2$-symmetric lattice simulations of real scalar fields give identical results (in terms of average amplitudes and energies) to those for $U(1)$-symmetric complex fields in the unitary gauge, since there exists only one physical, real and positive-definite degree of freedom that follows the same averaged dynamics of a real scalar field in a $\mathds{Z}_2$-symmetric potential.     
\end{itemize}

\section{Numerical lattice simulations} \label{app:lattice_simulations}

Our numerical analysis of the Higgs dynamics in its effective potential is carried out using the \texttt{$\mathcal{C}osmo\mathcal{L}attice$} code \cite{Figueroa:2021yhd, Figueroa:2016wxr} following the general procedure introduced in previous works \cite{Laverda:2023uqv, Laverda:2024qjt, Bettoni:2024ixe}. The first step consists in implementing the full effective potential in a model file that includes a curvature-dependent effective mass, the three-loop RGI running of $\lambda(\mu)$ in \eqref{eq:running_fit} with an explicit dependence on the top-quark mass and the additional higher-order operator in \eqref{eq:effective_potential_sextic}. The lattice parameters are chosen to guarantee the stability of the solution and the covering of all relevant scales. 

\begin{itemize}
    \item The number of lattice points per dimension has been fixed to $N=960$ for the simulations shown in Figure \ref{fig:gw_sextic}, with IR and UV momenta being defined by $\kappa_{\rm IR}=2\pi/L$, $\kappa_{\rm UV}= \sqrt{3} N \kappa_{\rm IR} / 2$, with $\kappa\equiv k/\sqrt{6\xi}\mathcal{H}_{\rm kin}$. The lattice size is $L=N \, \delta x$ and $\delta x = 4\pi \, \nu / N$ is the dimension of a single lattice cell. From the discussion in \cite{Bettoni:2019dcw}, we set the smallest momentum in the tachyonic band with {$\kappa_{\text{IR}} = \mathcal{H}_{\rm kin}$, while the largest amplified momentum is set to be much smaller than the lattice's UV momentum, i.e. $\sqrt{4\nu^2 - 1}\mathcal{H}_{\rm kin} \ll \kappa_{\text{UV}}$}. 

    \item The time-step is chosen according to the stability constraint $\delta t / \delta x \ll 1/\sqrt{d}$ \cite{Figueroa:2021yhd}, with $d=3$ the number of spatial dimensions. The stiff background expansion is simply obtained by fixing the equation-of-state parameter to $w=1$.

    \item A symplectic 4th order Velocity-Verlet method is used to evolve the scalar field. The initial conditions for our 3+1 classical lattice simulations are set as $h(0)=h'(0)=0$, in agreement with the inflationary picture in \cite{Bettoni:2019dcw}. Fluctuations over this homogeneous background are included as Gaussian random fields, as done customarily for systems with short classicalisation timescales \cite{Bettoni:2021zhq}. The evolution is deterministic up to a randomised base seed that we choose to keep constant. Such a choice does not influence the evolution of the system, since it quickly looses memory of its initial state.

    \item The computation of the GW spectrum is obtained by solving the modified equation of motion for metric perturbations computed in \cite{Bettoni:2024ixe} that take into account the presence of non-minimal interactions within the spectator field approximation.
\end{itemize}

\begin{figure}[tb]
    \centering
        \includegraphics[width=0.47\textwidth]{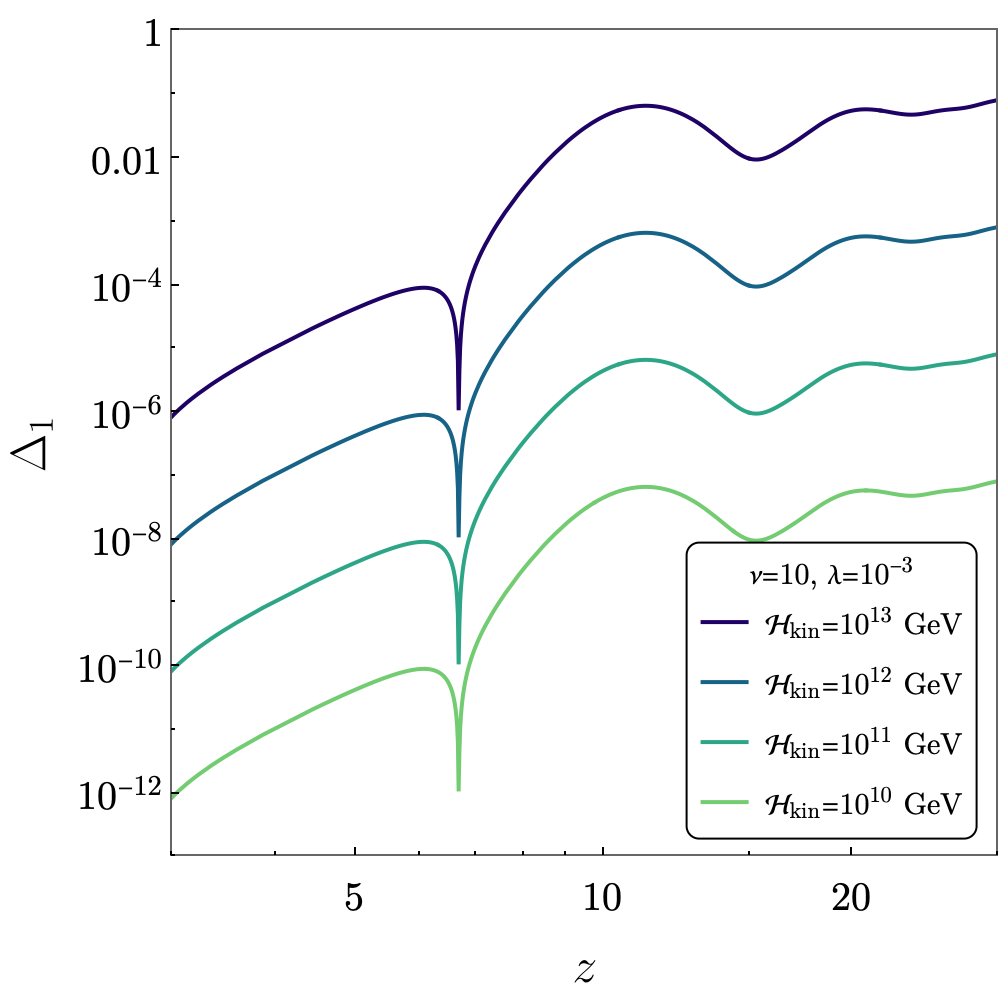}
        \hspace{1mm}
        \includegraphics[width=0.47\textwidth]{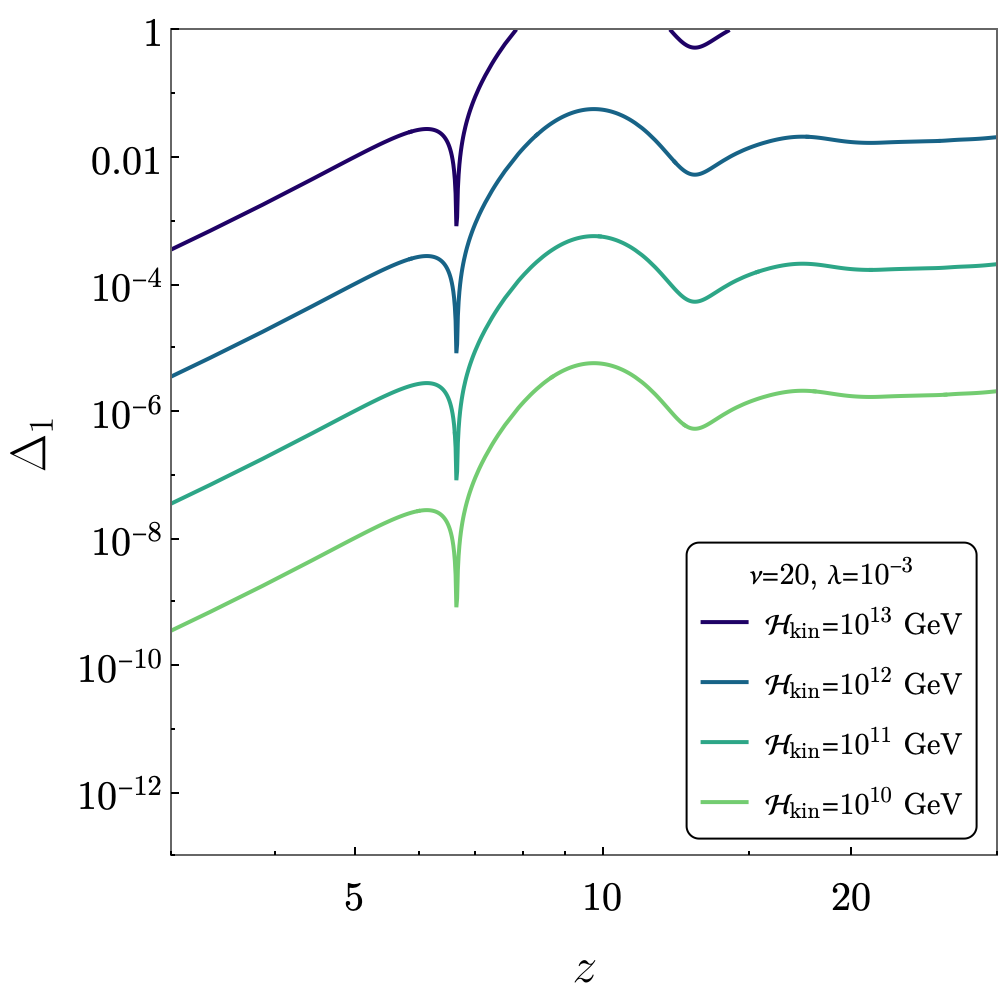}
        \hspace{4mm}
        \caption{Time evolution of the ratio $\Delta_1$ for benchmark scenarios with $\lambda=10^{-3}$ and $\nu=10, \; 20$ and different scales of kination $\mathcal{H}_{\rm kin}$.} \label{fig:delta_1}
\end{figure}

\begin{figure}[tb]
    \centering
        \includegraphics[width=0.47\textwidth]{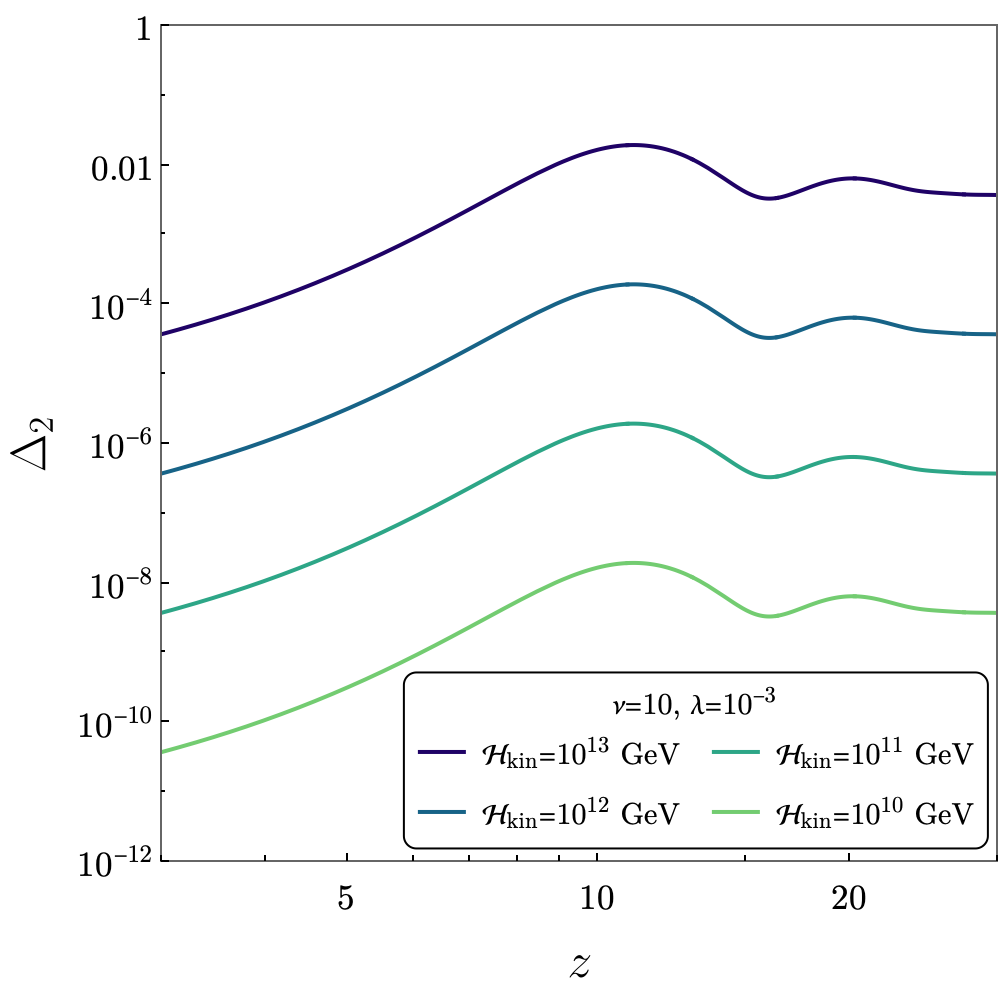}
        \hspace{1mm}
        \includegraphics[width=0.47\textwidth]{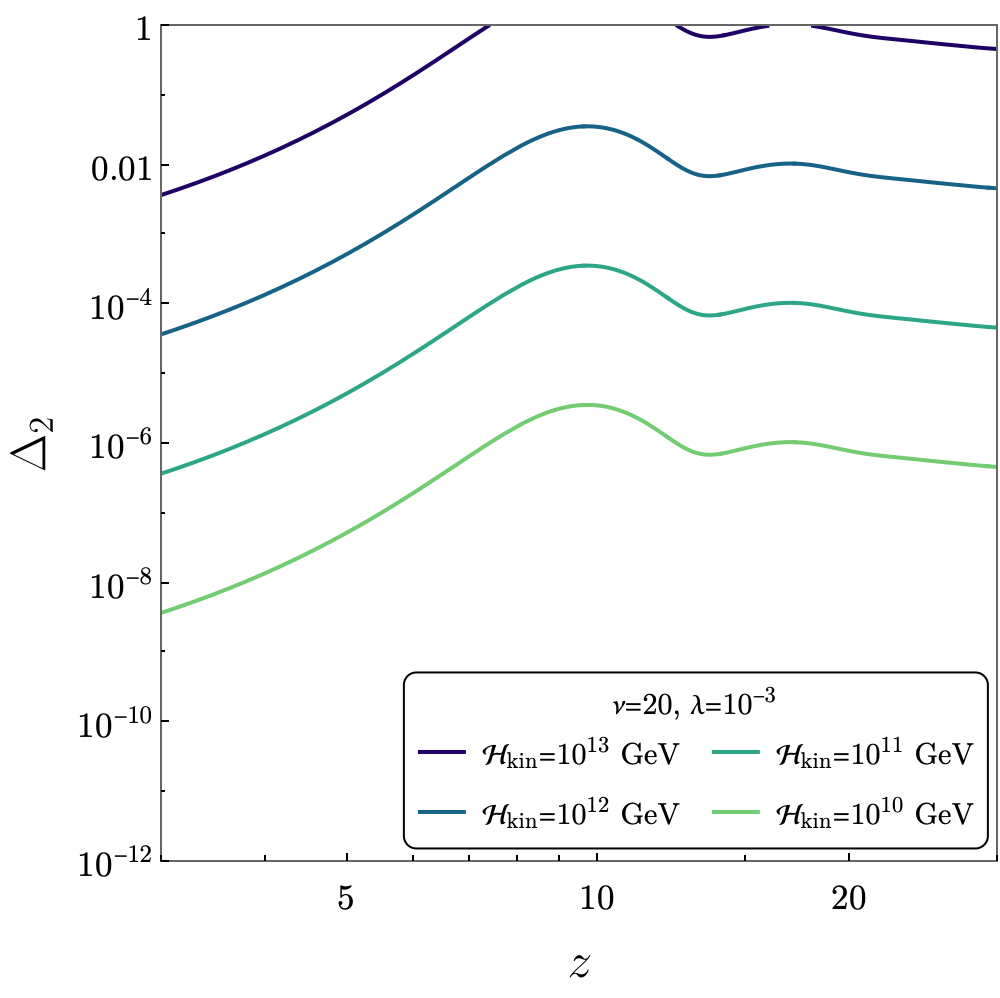}
        \hspace{4mm}
        \caption{Time evolution of the ratio $\Delta_2$ for benchmark scenarios with $\lambda=10^{-3}$ and $\nu=10, \; 20$ and different scales of kination $\mathcal{H}_{\rm kin}$.} \label{fig:delta_2}
\end{figure}

Following the discussion in Section \ref{sec:spectatorHiggs} and in the previous Appendix \ref{app:parametric_formulas}, we can asses the spectator field-regime thanks to some benchmark 3+1 lattice simulations. In particular, since we are interested in checking some general properties of the symmetry breaking, it is more instructive to fix the Higgs self-coupling to some constant value. Within the parameter space in Figure \ref{fig:Hkin_GW}, the scale-dependence of $\lambda$ leads to typical values $\lambda>10^{-3}$. Therefore, we perform a few lattice simulations with $\lambda=10^{-3}$, $\nu=10, \; 20$ and $N=256$ to probe the regime in which the spectator-field approximation is valid. Figure \ref{fig:delta_1} shows the time-evolution of the term $\Delta_1$ for different scales of kination. Crossing the value $\Delta_1=1$ implies a change in the sign of the Ricci scalar and the restoration of the unbroken phase in the effective potential. This happens generally when the tachyonic mass is large, i.e. for large scales of kination and large non-minimal couplings. Within the range $\mathcal{H}_{\rm kin} \lesssim 10^{12} \textrm{ GeV}$, the energy density of the spectator field is sufficiently smaller than the inflaton energy density not to lead to a faster symmetry restoration. 

The second consistency check of the spectator approximation is the smallness of $\Delta_2$, which is implied by imposing the cut-off scale of the effective theory on the Higgs fluctuations. In this case, we can make use of the parametric formula for $\langle h_{\rm br}^2 \rangle$ to estimate the constraint $\langle h_{\rm br}^2 \rangle \ll M_P^2/\xi^2$ in a parametric way. The result is already included in figure \ref{fig:Hkin_GW} and figure \ref{fig:res_cav_top_mass} as a light-blue-shaded area and it takes into account the running of the Higgs self-coupling. We maintained a conservative bound of $\langle h_{\rm br}^2 \rangle= 10^{-2} \times M_P^2/\xi^2$, thus accounting for the small discrepancy between the average amplitude at backreaction time and the largest average amplitude at the end of the first semi-oscillation. Figure \ref{fig:delta_2} shows the full evolution of $\Delta_2$ during the tachyonic phase and the beginning of the non-linear phase for a small number of benchmark points in our parameter space. Notice again, that the value of $\lambda=10^{-3}$ is chosen in a conservative way and that the typical value of the Higgs self-coupling in the vast majority of our parameter space is larger. The $1\%$ threshold excludes kination scales above $\sim10^{12} \textrm{ GeV}$, which we have taken as an upper limit for Figures \ref{fig:GW_absolute stability}, \ref{fig:GW_external_heating}, \ref{fig:mt_Hkin_instability_scale}.

\bibliographystyle{JHEP.bst}
\bibliography{bibliography}

\end{document}